\documentclass[letterpaper]{JHEP3}

\usepackage{amsmath,amssymb,bm}
\usepackage{latexsym}
\usepackage{cite}

\usepackage{epsf,epsfig}

\usepackage{graphicx}
\usepackage{graphicx, epsfig, amssymb} 
\usepackage{bm}
\usepackage{amsfonts}
\usepackage{amsmath}
\usepackage{mathrsfs}

\usepackage{amssymb}
\usepackage{graphicx}
\usepackage{epsfig}
\usepackage{amsfonts}

\usepackage{a4}
\usepackage{amsmath}

\usepackage{amsmath}
\DeclareMathOperator\arctanh{arctanh}

\newcommand{\ee}{\end{equation}}
\newcommand{\eea}{\end{eqnarray}}
\newcommand{\be}{\begin{equation}}
\newcommand{\bea}{\begin{eqnarray}}

\def\dalemb#1#2{{\vbox{\hrule height .#2pt
        \hbox{\vrule width.#2pt height#1pt \kern#1pt
                \vrule width.#2pt}
        \hrule height.#2pt}}}

\def\0{{\sst{(0)}}}
\def\1{{\sst{(1)}}}
\def\2{{\sst{(2)}}}
\def\3{{\sst{(3)}}}
\def\4{{\sst{(4)}}}
\def\5{{\sst{(5)}}}
\def\6{{\sst{(6)}}}
\def\7{{\sst{(7)}}}
\def\8{{\sst{(8)}}}

 \def\bd{\begin{document}} \def\ed{\end{document}}
\def\ds{\documentstyle} \let\fr=\frac \let\bl=\bigl \let\br=\bigr
\let\Br=\Bigr \let\Bl=\Bigl
\let\bm=\bibitem
\let\na=\nabla
\let\pa=\partial \let\ov=\overline
\def\fft#1#2{{#1 \over #2}}

\newcommand{\insertplot}[5]{\begin{figure}
 \hfill\hbox to 0.05in{\vbox to #5in{\vfill
 \inputplot{#1}{#4}{#5}}\hfill}
 \hfill\vspace{-.1in}
 \caption{#2}\label{#3}
 \end{figure}}

    \numberwithin{equation}{section}

\title{
Spinning black holes  in shift-symmetric Horndeski theory
}
\author{
Jorge F. M. Delgado, Carlos A. R. Herdeiro and Eugen Radu\\
Departamento de Matem\'atica da Universidade de Aveiro and \\
Centre for Research and Development  in Mathematics and Applications (CIDMA) \\
Campus de Santiago, 3810-183 Aveiro, Portugal}
\abstract{%
We construct spinning black holes (BHs) in shift-symmetric Horndeski theory. 
This is an Einstein-scalar-Gauss-Bonnet model wherein the (real) scalar 
field couples linearly to the Gauss-Bonnet curvature squared combination. The BH solutions constructed are  stationary, axially symmetric and asymptotically flat. They possess a non-trivial scalar field outside their regular event horizon; thus they have scalar hair. The scalar ``charge" is not, however, an independent macroscopic degree of freedom. It is  proportional to the Hawking temperature, as in the static limit, wherein the BHs reduce to the spherical solutions found by Sotirou and Zhou. The spinning BHs herein are found by  solving non-perturbatively the field equations, numerically.  We present an overview
of the parameter space of the solutions together with a study of their basic geometric and phenomenological properties.  These solutions are compared with the spinning BHs in the Einstein-dilaton-Gauss-Bonnet model and the Kerr BH of vacuum General Relativity. As for the former, and in contrast with the latter, there is a minimal BH size and small violations of the Kerr bound. Phenomenological differences with respect to either the former or the latter, however, are small for illustrative observables, being of the order of a few percent, at most. 
}
\keywords{Black holes, Horndeski, modified gravity}

\begin{document}


\section{Introduction }
Scalar-tensor theories of gravity have attracted much attention since the pioneering example of Brans-Dicke theory~\cite{Brans:1961sx}. The physical relevance of such models could be tested, in particular, in strong gravity systems, namely black holes (BHs). On the one hand, as it turns out, the BH solutions in Brans-Dicke theory, as well as in a large class of models where the scalar field is non-minimally coupled to the Ricci scalar, are the same as in General Relativity (GR)~\cite{Hawking:1972qk,Sotiriou:2011dz}. On the other hand, BHs in extended scalar-tensor models, namely those with higher curvature corrections are, generically, different from those of GR~\cite{Herdeiro:2015waa}. 

Within the class of scalar-tensor theories that possess higher curvature corrections, those including a real scalar field, $\phi$, with a canonical kinetic term, non-minimally coupled to the Gauss-Bonnet (GB) quadratic curvature invariant,
\begin{eqnarray}  
R^2_{\rm GB} \equiv  R_{\alpha\beta\mu\nu} R^{\alpha\beta\mu\nu} - 4 R_{\mu\nu} R^{\mu\nu} + R^2 \ ,
\end{eqnarray} 
 have attracted considerable interest. This is the class of \textit{Einstein-scalar-GB} (EsGB) models described by the action
\begin{eqnarray}  
\label{action}
\mathcal{S}=
\int d^4x \sqrt{-g} \left[  R - \frac{1}{2}
 \partial_\mu \phi\partial^\mu \phi
 + \alpha  f(\phi) R^2_{\rm GB}   \right], 
\end{eqnarray} 
where  
$\alpha $ is a dimensionful coupling constant and 
$f(\phi)$ is a dimensionless coupling function. In these models, the GB term becomes dynamical in four spacetime dimensions, and the equations of motion remain second order, which is typically not the case when higher curvature corrections are included in the action. Moreover, the GB term as a higher order correction is suggested from string theory~\cite{Zwiebach:1985uq}.

The status of BHs in the family of models~\eqref{action} depends on the properties of $f(\phi)$; its choice determines if $\phi=0$ is a consistent truncation of the equations of motion. There are two generic cases. Following the classification in~\cite{Astefanesei:2019pfq} for a cousin model, we call models where $\phi=0$ is \textit{not} a consistent truncation of the equations of motion {\it class I or dilatonic-type.}
In this class of EsGB models $\phi \equiv 0$ does $not$ solve the field equations.
 Thus the Schwarzschild/Kerr BH is not a solution. In terms of the coupling function, this class of models obeys (from the scalar field equation (\ref{KG-eq}) below)
\begin{eqnarray}
\label{condx}
f_{,\phi}(0)\equiv \frac{d f (\phi)}{d \phi}\Big |_{\phi=0} \neq 0\ .
\end{eqnarray}
A representative example of coupling for this class is the standard dilatonic coupling,
$
f (\phi)=e^{\gamma \phi}$, which emerges in Kaluza-Klein theory, string theory and supergravity. In this case $\phi$ is often referred to as the \textit{dilaton} field. BHs in the Einstein-dilaton-GB model were constructed in~\cite{Kanti:1995vq,Kleihaus:2015aje,Kleihaus:2011tg}, where they were shown to have a qualitatively novel feature: a minimal BH size, determined by the coupling constant $\alpha$. Some of these BHs are perturbatively stable~\cite{Kanti:1997br} and aspects of their phenomenology has been considered in $e.g.$~\cite{Cunha:2016wzk,Zhang:2017unx,Blazquez-Salcedo:2017txk}.

Models where $\phi=0$ is a consistent truncation are called 
{\it  class II or scalarised-type}.  
In this case  $\phi \equiv 0$ solves the field equations
and thus
 Schwarzschild and Kerr BHs are solutions of the full model. 
This demands that 
\begin{equation}
f_{,\phi}(0)\equiv \frac{d f (\phi)}{d \phi}\Big |_{\phi=0}= 0 \ .
\label{typeii}
\end{equation}
This condition holds, for instance, if one requires the model to be $\mathbb{Z}_2$-invariant under $\phi\rightarrow -\phi$. The Schwarzschild/Kerr BH solution is not, in general,  unique. 
These EsGB models may contain a second set of BH solutions, 
with a nontrivial scalar field profile -- {\it the scalarised BHs}. 
Such second set of BH solutions may, or may not, continuously connect with GR BHs.  Models within this class have been recenly under scrutiny in relation to BH spontaneous scalarisation - see $e.g.$~\cite{Doneva:2017bvd,Silva:2017uqg,Antoniou:2017acq,Cunha:2019dwb,Collodel:2019kkx}.
Two reference examples of coupling functions in this case are $f_1(\phi)=\gamma \phi^2$ and 
$
f _2(\phi)=e^{\gamma \phi^2} \ .
$
Although $f_1$ is the linearisation of $f_2$ (the constant term is irrelevant here) these two models have qualitatively different properties. Namely, the spherical scalarised BHs with the former coupling function are unstable against perturbations; but the ones with the latter coupling function can be stable~\cite{Blazquez-Salcedo:2018jnn}. 

In this paper we are interested in a model of class I, the linear coupling or \textit{shift symmetric} model. The coupling function is
\begin{eqnarray}
f(\phi)= \phi~,
\end{eqnarray} 
which implies the existence of a shift symmetry:  the equations of motion are  invariant under the transformation
\begin{eqnarray}
\label{shift-symm}
\phi \to \phi+\phi_0~,
\end{eqnarray} 
with $\phi_0$ an arbitrary constant.
This invariance results from the fact that 
in four spacetime dimensions 
 the GB term alone 
 is a total divergence. BHs in the model~\eqref{action} with~\eqref{shift-symm} have been first discussed by Sotiriou and Zhou (SZ)~\cite{Sotiriou:2014pfa,Sotiriou:2013qea}. This model falls within the Horndeski class~\cite{Horndeski:1974wa}, for which a no-scalar-hair theorem had been established~\cite{Hui:2012qt}. However, the SZ solution circumvents this theorem, since one of the assumptions (finitness of a certain current) is violated. The SZ solution has a minimal size, such as the BHs in Einstein-dilaton-GB. In fact, the model~\eqref{action} with~\eqref{shift-symm} can be seen as a linearisation of the Einstein-dilaton-GB model, and thus one expects similar properties for the BH solutions of both models. However, as pointed out above, models with a certain coupling function and its linearisation may have different properties. It has also been argued that the SZ could emerge dynamically in a gravitational collapse scenario~\cite{Benkel:2016rlz}.

The goal of this paper is to construct and study the basic physical properties of the spinning generalisation of the SZ solution, which, up to now, have not been considered. Astrophysical BHs have angular momentum. Thus, considering spinning BHs is fundamental to assess the physical plausibility of any BH model. This is, however, technically more challenging than for spherical BHs, in particular in the presence of higher curvature corrections, such as the GB invariant, as described below. 

This paper is organised as follows. In Section~\ref{sec2} we briefly discuss the equations of motion and some relevant properties of the model. In Section~\ref{sec3} we provide a short review of the spherical SZ solutions, as a warm up for the spinning case. In Section~\ref{sec4} we introduce the framework for the construction of spinning BHs, discussing the ansatz, boundary conditions, the physical quantities of interest and the numerical procedure. In Section~\ref{sec5} we describe the spinning BH solutions, its domain of existence, and the behaviour of different physical quantities. In Section~\ref{sec6} we present conclusions and remarks. Two appendices give some technical details on the construction of perturbative and extremal solutions.

\section{The model }
\label{sec2}

We consider a general EsGB model with the action~\eqref{action}. We use units such that $c=1=16\pi G$. Observe that the coupling constant has physical dimension  $[\alpha] \sim [L]^2$, where $L$ represents ``length".
Varying the action~ (\ref{action}) with respect to the metric tensor
 $g_{\mu\nu}$, 
we obtain 
the Einstein field equations
\begin{eqnarray}
\label{EGB-eq}
E_{\mu\nu}\equiv R_{\mu\nu}-\frac{1}{2}g_{\mu\nu} R -\frac{1}{2}T_{\mu\nu} =0\ .
\end{eqnarray}
The {\it effective} energy-momentum tensor has two distinct components,
 \begin{eqnarray}
\label{Teff}
 T_{\mu\nu} = T_{\mu\nu}^{(s)}-2\alpha T_{ \mu\nu}^{(GB)} \ .
\end{eqnarray}
The first one is due to the scalar kinetic term in (\ref{action})
\begin{eqnarray}
T_{\mu\nu}^{(s)}=\partial_{\mu} \phi\partial_{\nu} \phi -\frac12 g_{\mu\nu}\partial_\alpha\phi\partial^\alpha\phi \ ;
 \end{eqnarray} 
the second one is due to the scalar-GB  term in (\ref{action}), and reads
 \begin{eqnarray}
\label{Teff2}
 T_{\mu\nu}^{(GB)}=   P_{\mu\gamma \nu \alpha}\nabla^\alpha \nabla^\gamma f(\phi) \ ,
 \end{eqnarray}  
where we have defined
 \begin{eqnarray}
		P_{\alpha\beta\mu\nu} & \equiv & -\frac14 \varepsilon_{\alpha\beta\rho\sigma} R^{\rho\sigma\gamma\delta} \varepsilon_{\mu\nu\gamma\delta} \\
		  &= & R_{\alpha\beta\mu\nu}+ g_{\alpha\nu} R_{\beta\mu} - g_{\alpha\mu} R_{\beta\nu} + g_{\beta\mu} R_{\alpha\nu}-g_{\beta\nu} R_{\alpha\mu} 
			+\frac12 \left( g_{\alpha\mu}g_{\beta\nu} - g_{\alpha\nu}g_{\beta\mu}\right) R \ .
\nonumber
 \end{eqnarray}   
Here, $ \varepsilon_{\alpha\beta\rho\sigma}$ is the Levi-Civita tensor. The equation for the scalar field is 
\begin{eqnarray}
\label{KG-eq}
\Box \phi +\alpha  \frac{d f (\phi)}{d \phi} R^2_{\rm GB}
 =0 \ .
\end{eqnarray}

As pointed out in the introduction, the GB term is a total divergence:
\begin{eqnarray}
\label{totder}
 R^2_{\rm GB} =\nabla_\mu P^\mu \ ,
\end{eqnarray} 
where the vector $P^\mu$ takes a particularly simple form 
\cite{Yale:2010jy}
for a spacetime
  possessing a Killing vector $\partial/\partial t$ 
($t$ is the time coordinate),
\begin{eqnarray}
 P^\mu=4 P_{\nu}^{~\alpha \mu t} \Gamma^\nu_{t \alpha} \ .
\end{eqnarray} 
Thus  the  transformation 
(\ref{shift-symm})
does not change the equations of the model.
Moreover, (\ref{totder})
implies that the equation for the scalar field (\ref{KG-eq})
can be written as
\begin{eqnarray}
\label{relD}
\nabla_\mu J^\mu=0\ , \qquad {\rm with}~~
J^\mu=  \partial^\mu \phi+\alpha P^\mu \ .
\end{eqnarray} 
As we shall see, 
a consequence of this relation is that
the scalar `charge' 
(as read off from the asymptotically leading monopolar mode) is just the Hawking temperature
of BH~\cite{Prabhu:2018aun}.

In this work we shall be interested in stationary, axially symmetric  solutions. They possess two asymptotically measured global charges:
the mass $M$ and the angular momentum $J$.
There is also a scalar charge $Q_s$, but it is not an independent quantity; it depends on the BH mass and angular momentum.
Thus the scalar hair is of secondary type~\cite{Herdeiro:2015waa}. 
 Also,
note that
the shift symmetry
(\ref{shift-symm})
 is broken  
  by imposing $\phi(\infty)=0$. Horizon quantities of physical interest, on the other hand, include
the Hawking temperature $T_H$,
the horizon area $A_H$
and the entropy $S$,
whose concrete expressions   are given below.

Since the equations of the model are invariant under the transformation
\begin{eqnarray}
\label{scale}
r\to \lambda r\ , \qquad \alpha \to \lambda \alpha \ ,
\end{eqnarray} 
where $\lambda>0$ is an arbitrary constant, the most meaningful physical quantities must be invariant under (\ref{scale}).
Considering how the 
various global quantities transform under this scaling
($e.g.$ $M\to \lambda M$, $J\to \lambda^2 J$, $etc.$) we normalise the various quantities 
$w.r.t.$ the mass of the solutions. 
In this way, we define the \textit{reduced} 
angular momentum $j$,
horizon area $a_H$,
entropy $s$
and
Hawking temperature $t_H$
as
\begin{eqnarray}
\label{scale2}
j\equiv \frac{J}{M^2}\ , \qquad 
a_H\equiv \frac{A_H}{16\pi M^2}\ , \qquad s\equiv \frac{S}{4\pi M^2} \ , \qquad 
t_H\equiv 8 \pi T_H M \ .
\end{eqnarray} 
Alternatively, one can
define dimensionless reduced variables $w.r.t.$ the coupling constant $\alpha$
(we recall that  $[\alpha] \sim [L]^2$). 

 \section{Spherically symmetric black holes}
\label{sec3}

Before discussing the case of spinning BHs,
it is of interest to  review the construction and basic properties 
of the static, spherically symmetric BHs, 
the SZ solutions~\cite{Sotiriou:2014pfa,Sotiriou:2013qea}.
As we shall see, they contain valuable information, and share some key properties
with their rotating counterparts, being easier to study since 
they are found by solving a set of ordinary differential equations.
Moreover, a perturbative {\it exact} solution is  available in the static case,
which is discussed in Appendix~\ref{a1}.

 \subsection{The equations and boundary conditions}
The spherical BHs of~\eqref{action} with~\eqref{shift-symm}  can be found using 
Schwarzschild-like coordinates, with a metric ansatz containing two unknown functions,
 \begin{equation}
\label{s1}
 ds^2=-N(r) \sigma^2(r) dt^2+\frac{dr^2}{N(r)}+r^2 d\Omega_2^2\ , \qquad {\rm with} \ \ \ N(r)\equiv 1-\frac{2m(r)}{r}\ ,
\end{equation}
where $r$ and $t$ are the radial and time coordinate, respectively, 
$d\Omega_2^2$ is the metric on the unit round $S^2$ and
 $m(r)$ is the Misner-Sharp  mass \cite{Misner:1964je},
which obeys $m(r)\to M$ as $r\to \infty$.
The scalar field $\phi$ 
is a function of $r$ only. The Schwarzschild
BH corresponds to $\phi=0$, $m(r)=r_H/2=$constant,
$\sigma(r)=1$. One can easily verify that for
$\alpha\neq 0$
 this is not a solution of the model in this work.

The advantage of this metric gauge choice is the simple form of the Einstein equations (\ref{EGB-eq}), which yield the
 generic relations
 \begin{equation}
m'=-\frac{r^2}{4} T_t^{t }\ , \qquad \frac{\sigma'}{\sigma}=\frac{r}{4 N} ( T_r^{r }-T_t^{t }) \ .
\label{ss143}
\end{equation}
For the considered EsGB model, 
the diagonal components of the effective energy-momentum tensor contain second derivatives of 
the metric functions
$N,\sigma$.
However, 
one can  find a
 suitable combination of the field  equations
such that
the functions $m,\sigma$  
still solve first order equations. These equations are
\begin{eqnarray}
\label{eq-N}
&&
\left[ 
         1+2\alpha (1-3N)\frac{\phi'}{r} 
\right] m'
-
\left\{
\frac{N}{8}r^2 \phi'^2+\alpha(1-N)\left[(1-3N)\frac{\phi'}{r}+2N\phi''\right]
\right\}
=0 \ ,
\\
&&
\label{eqs-spherical}
 \frac{\sigma'}{\sigma}
\left[1+2 \alpha (1-3 N)\frac{\phi'}{r}\right]
-\frac{1}{4 r} 
 \left[ r^2\phi'^2 +8\alpha (1-N) \phi'' \right]
=0 \ .
\end{eqnarray}
The Einstein equations contain also a second order equation which provides a constraint, being a
linear combination of (\ref{eq-N}) and (\ref{eqs-spherical}) 
together with their first derivatives.

The scalar field $\phi$ is a solution of a 2nd order equation 
in terms of $N$ and $\phi'$
only
\begin{eqnarray}
\nonumber
&&
\phi''
\bigg[
 1+\frac{2\alpha}{r}(1-7N)\phi'
-\frac{24 \alpha^2}{r^4}
\left[
2(1-N)^2+r^2N(1-3N)\phi'^2
\right]
+\frac{8 \alpha^3 N\phi'}{r^5}
\big[
24(1-N)^2
\\
\nonumber
&&
{~~~}
+r^2\{1+3N(2-5N)\}\phi'^2
\big]
\bigg]
+\frac{1}{r}
\bigg[
\left(1+\frac{1}{N}\right)\phi'
+\frac{2\alpha}{r^3 N}
\bigg[
  6(1-N)^2+r^2(1-N-12N^2)\phi'^2
	\\
\nonumber
&&
{~~~}
	  -\frac{1}{8}r^4 N^2 \phi'^4
\bigg]
-\frac{8\alpha^2 \phi'}{r^4}
  \bigg[
	       6(1+N^2)-r^2\phi'^2(1+21N^2)
				-N\left(12-10r^2\phi'^2+\frac{1}{8}r^4 \phi'^4\right)
	\bigg]
		\\
\label{eq-phi}
&&
{~~~~~~~~~~~~~~~~~~~~~~~~~~~~~~}
+\frac{8\alpha^3}{r^3} 
          (1-3N)^2(1-5N)\phi'^4
\bigg]=0 \ .
\end{eqnarray}
This approach leads to a good accuracy of the numerical results, 
and can easily be generalized for an arbitrary coupling function $f(\phi)$.

The approximate form of the solutions valid for large-$r$
reads
\begin{equation}
N(r)=1-\frac{2M}{r}+\frac{Q_s^2}{4r^2}+\dots\ , \qquad 
\sigma(r)=1-\frac{Q_s^2}{8r^2}+\dots\ , \qquad
\phi(r)=-\frac{Q_s}{r}- \frac{Q_sM}{r^2}+\dots \ ,
\end{equation}
in terms of mass $M$ and a scalar ``charge" $Q_s$.
Close to the event horizon, located at $r=r_H$,
the solutions possess an approximate expression  
as a power series in $r-r_H$, with
\begin{eqnarray}
\nonumber
N(r)&=&N_1(r-r_H)+\dots\ , \qquad
\sigma(r)=\sigma_H+\sigma_1(r-r_H)+\dots\ , \\
\phi(r)&=&\phi_H+\phi_1(r-r_H)+ \phi_2(r-r_H)^2+\dots\ ,
\end{eqnarray}
where
\begin{eqnarray}
 N_1=\frac{1}{2\alpha \phi_1+r_H}\ , \qquad \sigma_1=\frac{(16\alpha \phi_2+\phi_1^2 r_H^2)\sigma_H}{4(2\alpha \phi_1+r_H)} \ ,
\end{eqnarray}
while $\phi_2$ is a complicated function of $\phi_1$, $r_H$ and $\alpha$.
The Hawking temperature, horizon area and entropy of the  solutions,
as computed from the formalism in the next Section, are given by
\begin{eqnarray}
T_H=\frac{N_1 \sigma_H}{4\pi}\ , \qquad A_H=4\pi r_H^2\ ,  \qquad S=\pi r_H^2+4\pi \alpha \phi_H \ .
\end{eqnarray}
The field equations imply that
the first derivative of the scalar field,  $\phi_1$, is a solution of the quadratic equation
\begin{eqnarray}
\label{eqphi1}
\phi_1^2+\frac{ r_H}{2\alpha }\phi_1+\frac{6}{r_H^2}=0\ ,
\end{eqnarray}
which implies the following condition for 
the existence of a real root 
\begin{eqnarray}
\label{condi}
\frac{\alpha}{r_H^2}<\frac{1}{4\sqrt{6}}\simeq 0.10206~.
\end{eqnarray} 
This requirement translates into
the following coordinate independent condition
between the horizon size and the coupling constant
$\alpha$
\begin{eqnarray}
\label{cond}
A_H>16\pi \sqrt{6}\alpha\ .
\end{eqnarray}
We remark that $A_H=4\pi r_H^2$ for the metric ansatz employed here.
Thus, for a theory with a given value of the input parameter $\alpha>0$,
the BHs are not smoothly connected with the Minkowski vacuum.
There is minimal horizon size and a mass gap~\cite{Sotiriou:2014pfa,Sotiriou:2013qea}, just as for BHs in the Einstein-dilaton-GB model~\cite{Kanti:1995vq,Kleihaus:2015aje,Kleihaus:2011tg}.

 \subsection{The solutions}
The parameter space of solutions can be scanned by starting with the Schwarzschild
BH ($\alpha=0$) and increasing the value of $\alpha$ for fixed $r_H$. 
When
appropriately scaled,
they form a line, starting from the smooth GR limit
and ending at a \textit{critical} solution
where the condition 
(\ref{cond})
is violated, and 
where the maximal value of the ratio
$\alpha/M^2$ (around $0.32534$) is achived.
Once the critical  configuration is reached, the solutions cease to exist in the parameter space. Physically this means that the EsGB BHs have a minimal size and mass,
 for given $\alpha$. A possible interpretation is that the GB term provides a repulsive contribution, becoming overwhelming for sufficiently small BHs, thus preventing the existence of an event horizon. The full set of static solutions will be shown below in Fig.~\ref{dom0} (the blue dotted line with $j=0$)
as a function of the dimensionless parameter $\alpha/M^2$.

As discussed in Appendix~\ref{a1}, a simple perturbative solution can be found
as a power series in the parameter
\begin{eqnarray}
\label{beta}
 \beta\equiv \frac{\alpha}{r_H^2}=\frac{4\pi \alpha}{A_H} \ .
\end{eqnarray}
The results in Appendix~\ref{a1} imply the following expressions
\begin{eqnarray}
&&
a_H=\frac{A_H}{16\pi M^2}=
1-\frac{98 }{5  }\beta^2
+\frac{146378 }{1925 }\beta^4
-\frac{42468831605804  }{13266878625} \beta^6
+\dots\ ,
\\
\label{fin-res}
&&
t_H=8 \pi T_H M=1+\frac{146}{15  }\beta^2
+\frac{1410898 }{17325  }\beta^4
+\frac{72356439488}{57432375  }\beta^6
+\dots\ ,
\\
\nonumber
&&
s=\frac{S}{4 \pi M^2}=
 1+\frac{146}{15 }\beta^2
-\frac{13451026 }{51975  }\beta^4
+\frac{25584053312 }{57432375}\beta^6
+\dots\ ,
\\
\nonumber
&&
q=\frac{Q_s}{M}=
8\beta
-\frac{1184 }{15  }\beta^3
-\frac{4614784}{17325 }\beta^5
+\dots\ ,
\\
\nonumber
&&
\phi(r_H)= \frac{22}{3} \beta
+\frac{40516  }{675  }\beta^3
-\frac{7057522938136377682 }{119373478599375}\beta^7
+\dots. \ .
\end{eqnarray}
Interestingly,  all corrections to 
the reduced temperature $t_H$ are positive. That is, for the same mass, 
the shift symmetric Hordenski BH is `hotter'. 
For the other quantities, no clear generic pattern emerges.

We have found that the perturbative solution provides a
very good approximation to the numerical results. This follows from the smallness of the parameter $\beta$. In fact, condition
(\ref{condi}) implies  $\beta_{\rm max} \simeq 0.102062$.
As such, the contribution of the
higher order terms in $\beta$ quickly
becomes irrelevant.

 \section{Spinning black holes: the framework}
\label{sec4}
 \subsection{Ansatz and boundary conditions}

To obtain stationary and
axi-symmetric BH spacetimes, possessing
two commuting Killing vector fields, $\xi$ and $\eta$, we use a coordinate system adapted to these symmetries.
Then
$
\xi = \partial_t,
$
$
\eta=\partial_\varphi,
$
and we
 consider a metric  ansatz which has been employed in the past 
for the study of 
Kerr BHs with scalar hair~\cite{Herdeiro:2014goa}.
In terms of the spheroidal coordinates $r,~\theta$ and $\varphi$ (with $t$ the time coordinate), the
 metric line element reads: 
\begin{equation}
\label{ansatz}
ds^2=-e^{2F_0} N dt^2+e^{2F_1}\left(\frac{dr^2}{N }+r^2 d\theta^2\right)+e^{2F_2}r^2 \sin^2\theta (d\varphi-W dt)^2\ ,  \ \ \ \ N\equiv 1-\frac{r_H}{r}\ ,
\end{equation} 
where the metric functions
$F_i,W$, as well as the scalar field $\phi$,
depend on $r,\theta$ only and $r_H>0$ is an input parameter again describing the location of the event horizon.
The coordinates $\theta,\varphi$ and $t$
possess the  usual range, while $r_H\leqslant r <\infty$.
The vacuum Kerr BH
can be written in this form, the corresponding expressions of
$F_0,F_1,F_2$ and $W$ being displayed in Appendix A of~\cite{Herdeiro:2015gia}.

Finding BH solutions with this ansatz requires defining boundary behaviours. We have made the following choices. For the solutions to approach at spatial infinity ($r\rightarrow \infty$) a Minkowski spacetime we require
\begin{equation}
\lim_{r\rightarrow \infty}{F_i}=\lim_{r\rightarrow \infty}{W}=\lim_{r\rightarrow \infty}{\phi}=0\ .
\end{equation}
Since the scalar field is massless, one can construct an approximate solution of the field equations  
compatible with these asymptotics as a power series in $1/r$. The leading order terms of such an expansion
are:
\begin{eqnarray}
\label{r-infty}
\nonumber
&&
F_0(r,\theta)=\frac{c_t}{r} 
+\dots\ , \qquad
F_1(r,\theta)=-\frac{c_t}{r} 
+\dots\ , \qquad
F_2(r,\theta)=-\frac{c_t}{r} 
+\dots\ , \nonumber \\
&&
W(r,\theta)=\frac{c_\varphi}{r^3} 
+\dots\ , \qquad
\phi(r,\theta)=\frac{Q_s}{r  }+\dots \ ,
\end{eqnarray} 
where $c_t$, $c_\varphi$ and $Q_s$ are constant parameters to be fixed by the numerics.

Axial symmetry, together with regularity at the axis impose
the following boundary conditions on the symmetry axis, $i.e.$ at $\theta=0,\pi$:
\begin{equation}
\partial_\theta F_i = \partial_\theta W = \partial_\theta \phi = 0 \ .
\end{equation}
As before,  an approximate expansion of the solution compatible with these boundary conditions can be constructed;
as an illustration, at $\theta=0$ one finds
\begin{eqnarray}
\label{t0}
{\cal F}_a(r,\theta)= {\cal F}_{a0}(r)+\theta^2 {\cal F}_{a2}(r)+\mathcal{O}(\theta^4)\ ,
\end{eqnarray} 
where ${\cal F}_a =\{F_0, F_1, F_2, W; \phi\}$. The essential data, which is fixed by the numerics, is encoded in the 
functions ${\cal F}_{a0}=\{F_{i0},W_{0},\phi_{0}\}$.  
Moreover, the absence of conical singularities implies also that 
%
$
F_1=F_2 
$
on the symmetry axis.
Focusing on BHs with parity reflection symmetry,
we need to consider the solutions only
for $0 \leqslant \theta \leqslant \pi/2$.  
	Then, the functions
$F_i,~W$ and $\phi$ 
satisfy the following boundary conditions on the equatorial plane ($\theta=\pi/2$)
\begin{equation}
\partial_\theta F_i\big|_{\theta=\pi/2} = \partial_\theta W\big|_{\theta=\pi/2} =\partial_\theta \phi\big|_{\theta=\pi/2} = 0 \ .
\end{equation}

For the metric ansatz~\eqref{ansatz}, the event horizon is located at a surface with constant radial variable, $r=r_H>0$.
By introducing a new radial coordinate 
\begin{equation}
x=\sqrt{r^2-r_H^2} \ ,
\label{x}
\end{equation}
the horizon boundary conditions and numerical treatment of the problem simplify. These boundary conditions are 
\begin{equation}
\partial_x F_i \big|_{x=0}= \partial_x \phi  \big|_{x=0} =  0\ , \qquad W \big|_{x=0}=\Omega_H\ ,
\label{bch1}
\end{equation}
where $\Omega_H $ is the horizon angular velocity, and 
the Killing vector $\chi =\xi+\Omega_H \eta$ is orthogonal and null on the horizon.
These conditions are consistent with the near horizon solution 
\begin{eqnarray}
\label{rh}
{\cal F}_a(r,\theta)= {\cal F}_{a0}(\theta)+x^2 {\cal F}_{a2}(\theta)+\mathcal{O}(x^4)\ ,
\end{eqnarray}  
where the essential functions are
${\cal F}_{i0}$  
(also $F_0\big |_{r_H}=F_1\big |_{r_H}$).

 \subsection{Quantities of interest and a Smarr relation}

Many quantities of interest are 
encoded in the metric functions at the horizon or at infinity.
Considering first horizon quantities. The
Hawking temperature is $T_H={\kappa}/({2\pi})$, where $\kappa$ is the surface gravity
defined as $\kappa^2=-\frac{1}{2}(\nabla_a \chi_b)(\nabla^a \chi^b)|_{r_H}$, 
and the event horizon area $A_H$. 
These are computed as
\begin{eqnarray}
\label{THAH}
&&
T_H=\frac{1}{4\pi r_H}e^{F_0(r_H,\theta)-F_1(r_H,\theta)} \ ,
\qquad 
A_H=2\pi r_H^2 \int_0^\pi d\theta \sin \theta~e^{F_1(r_H,\theta)+F_2(r_H,\theta)} \ .
\end{eqnarray}
The horizon angular velocity $\Omega_H$ is fixed by the horizon value of the metric function $W$,
\begin{eqnarray}
\label{OmegaH}
\Omega_H=-\frac{g_{\varphi t}}{g_{tt}}\bigg|_{r_H}=W \bigg|_{r_H}.
\end{eqnarray}

The total (ADM)  mass $M$ and angular momentum $J$ of the BHs
are read off from the asymptotics of $g_{tt}$ and $g_{\varphi t}$,
\begin{eqnarray}
\label{asym}
g_{tt} =-1+\frac{2GM}{r}+\dots \ , \qquad ~~g_{\varphi t}=-\frac{2GJ}{r}\sin^2\theta+\dots \ .
\end{eqnarray}
These global quantities can be split into the horizon and bulk contributions - see, $e.g.$,~\cite{Townsend:1997ku}.
These are, respectively $M_H$ and $J_H$, computed as a Komar integrals on the horizon, and $M_\phi$ and $J_\phi$,
computed as volume integrals of the appropriate {\it effective} 
energy-momentum tensor components:
\begin{eqnarray}
\label{TotalMass}
&&
M = M_H+M_\phi\ , \qquad \qquad 
 M_\phi\equiv -2\int_\Sigma dS_\mu\bigg( T_{\nu}^{\ \mu}  \xi^{\nu}-\frac{1}{2}T \xi^\mu \bigg)\ ,
\\
\label{TotalAngularMomentum}
 &&
J = J_H +J_\phi\ , \qquad \qquad
 J_\phi\equiv \int_\Sigma dS_\mu \left( T_{\nu}^{\mu}  \eta^{\nu} -\frac{1}{2}T  \eta^{\mu} \right) \ ,
\end{eqnarray}
where $\Sigma$ is a spacelike surface, bounded by the 2-sphere at infinity
$S^2_\infty$ and the spatial section of the horizon $H$. $M_\phi$ and $J_\phi$
encode the contribution of the \textit{effective} ``matter"  distribution to the total
mass and angular momentum.
For Kerr BHs, $M=M_H$ and $J=J_H$;  
this is not so for EsGB BHs.  
Moreover,
since $T_t^{t(\phi)}-\frac{1}{2}T^{(\phi)}=T_\varphi^{t(\phi)}=0$,
only the GB part of the \textit{effective} energy-momentum tensor (\ref{Teff})
contributes to the 
energy and angular momentum ``matter" densities.

The solutions can be shown to obey the  Smarr-type law
\begin{eqnarray}
\label{smarr}
M +2\Omega_H J+ M_s=2 T_H S\ ,
\end{eqnarray}
where $S$ is the entropy as computed from Wald's formula 
\cite{Wald:1993nt},
\begin{eqnarray}
\label{S}
S=S_E+S_{sGB}\ , \qquad 
S_E=\frac{A_H}{4}\ , \qquad  S_{sGB}=\frac{\alpha}{2} \int_{H} d^2 x \sqrt{h}\phi {\rm  R} \ ,
\end{eqnarray}
and 
${\rm  R}$ is the Ricci scalar of the induced horizon metric $h$.
In the Smarr-type law, $M_s$ is a contribution of the scalar field
\begin{eqnarray}
\label{sup}
 M_s= \frac{1}{2} \int_\Sigma d^3x \sqrt{-g} \partial_\mu \phi\partial^\mu \phi \ ,
\end{eqnarray}
 which can also be expressed as an integral of $\phi R^2_{\rm GB}$ term.

Also, by integrating (\ref{relD})
over an hypersurface bounded by the event horizon and
the sphere at infinity  
one can prove the following 
 interesting relation
\begin{eqnarray}
\label{Qs}
Q_s=16 \pi \alpha T_H\ .
\end{eqnarray}  
This proportionality between 
the scalar charge 
and the Hawking temperature is a unique feature of the
shift symmetric 
EsGB model, see the discussion in \cite{Prabhu:2018aun}.

The EsGB BHs satisfy also 
the first law  
\begin{eqnarray}
\label{first-law}
dM=T_H dS +\Omega_H dJ \  .
\end{eqnarray}

\subsection{The numerical approach}

In our approach, the field equations reduce to a set of five 
coupled non-linear elliptic partial differential equations for the functions 
${\cal F}_a =(F_0, F_1, F_2, W; \phi)$,
which are found by plugging the ansatz 
(\ref{ansatz}) together with $\phi=\phi(r,\theta)$ 
into the field eqs.~(\ref{EGB-eq}), (\ref{KG-eq}).
They  consist   of 
the Klein-Gordon equation (\ref{KG-eq})
together with suitable combinations of the Eintein equations (\ref{EGB-eq}) 
$
\{ 
E_r^r+E_\theta^\theta=0;~
E_\varphi^\varphi=0;~
E_t^t=0;~
E_\varphi^t=0
 \}.
$ 
The explicit form of the equations solved in practice is too complicated to display here; 
each equation containing around 250 independent terms.
Also,  the remaining equations 
$E_\theta^r =0$ 
and
$E_r^r-E_\theta^\theta  =0$ 
are not solved directly, they
yielding two constraints  which are monitored in numerics. Typically they are satisfied at the level of the overall numerical accuracy. We remark that one can 
verify that
the remaining equations vanish identically,
$E_r^\varphi =E_r^t =E_\theta^\varphi =E_\theta^t =0$,
the circularity condition being satisfied.
As such, the employed ansatz is consistent,
a fact which is not \textit{a priori} guaranteed (see~\cite{VanAelst:2019kku}
for a discussion in an Einstein-scalar field model which leads 
to a non-circular metric form).

Our numerical treatment can be summarised as follows.
We restrict the domain of integration to the region outside the horizon. Then, 
the first step is to introduce the new radial variable 
$\bar x=x/(1+x)$ 
which maps the semi--infinite region $[0,\infty)$ to the finite region $[0,1]$, where  $x$ is given by~\eqref{x} and $r$ is the radial variable in 
the line element 
(\ref{ansatz}).
Next, the equations for ${\cal F}_a$
are discretised on a grid in $\bar x$ and $\theta$. 
 Most of the results in this work have been found for  
 an equidistant grid with $300 \times 40$ points. 
The grid covers the integration region
$0\leqslant \bar x \leqslant 1$ and $0\leqslant \theta \leqslant \pi/2$. 

The equations for ${\cal F}_a$
  have been solved  subject to the boundary conditions 
 introduced above.  
All numerical calculations  
are performed by using a professional package \cite{schoen},
which employs a  Newton-Raphson method.  
This code uses  the finite difference method, providing also an error estimate for each unknown function.
For the solutions in this work,
the maximal numerical error 
for the functions is estimated to be on the order of $10^{-3}$. 
The Smarr relation (\ref{smarr}) 
provides a further test  of the numerical accuracy, leading to error estimates of the same order. 

In our numerical scheme, there are three input parameters: 
${\bf i)}$  the event horizon radius $r_H$;
 ${\bf ii)}$
the event horizon angular velocity $\Omega_H$
 in the metric ansatz (\ref{ansatz})
and
 ${\bf iii)}$
 the coupling constant  $\alpha$ in the action (\ref{action}).
The quantities of interest are computed from the numerical output.
 For example, the mass $M$,  and the angular momentum $J$
are extracted from the  asymptotic expressions (\ref{asym}),
while the Hawking temperature, the entropy and the horizon area 
are obtained from the event horizon data.

The results
reported in this work are obtained from around twenty thousand solution points. 
For all these BHs we
have monitored the Ricci and the Kretschmann scalars, 
and, at the level of the numerical accuracy, we have
not observed any sign of a singular behaviour on and outside the horizon
(see, however, the discussion below on the limiting solutions).

 \section{Spinning black holes: numerical results}
\label{sec5}

 \subsection{General properties and limiting behaviour}

In an approach based on the Newton-Raphson method   
a good initial guess for the profile of the various functions is an essential condition for a successful implementation.
The spinning solutions in this work can be constructed by using 
two different
routes. 
In the first approach, one uses the profile of a Kerr BH with given $r_H,\Omega_H$
 as
an initial guess for EsGB solutions\footnote{We mention that, similar to the static limit, the 
scalar field equation (\ref{KG-eq}) possesses a nontrivial solution in a fixed Kerr background,
which inherits most of the basic properties of the backreacting generalization.
In particular, the  scalar charge-Hawking temperature relation (\ref{Qs}) holds also in this case,
while the scalar field
appears to diverge as the extremal Kerr limit is approached.} with a small value of the ratio
$\alpha/r_H^2$.
The iterations
converge and, repeating the procedure, one obtains in this way solutions with large $\alpha$.
In the second approach, one starts instead with  spherically symmetric solutions of EsGB, either obtained numerically or from the perturbative expansion. These can  also be studied within the ansatz (\ref{ansatz}),
with $W=0$, $F_i$ being functions of $r$ only and with
$F_1=F_2$. Then, starting with an EsGB spherical BH with a given $r_H$ and $\alpha \neq 0$, rotation is introduced by introducing and slowly increasing  $\Omega_H$.

For all solutions we have found, the metric functions ${\cal F}_a$, together with their first and second derivatives with respect
to both $r$ and $\theta$ have smooth profiles. This leads to finite curvature invariants on the full domain of integration,
in particular at the event horizon. 
The shape of the metric functions $F_0,F_1,F_2$ and $W$ is similar to those  in the $\alpha = 0$ case. 
The maximal deviation from the Einstein gravity profiles (with the same 
input parameters $r_H,\Omega_H$)
is near the horizon.
At the same time, the scalar field may possess a complicated angular dependence,
in particular for fast spinning configurations.

\begin{figure}[t!]
\begin{center}
\includegraphics[height=.255\textheight, angle =0]{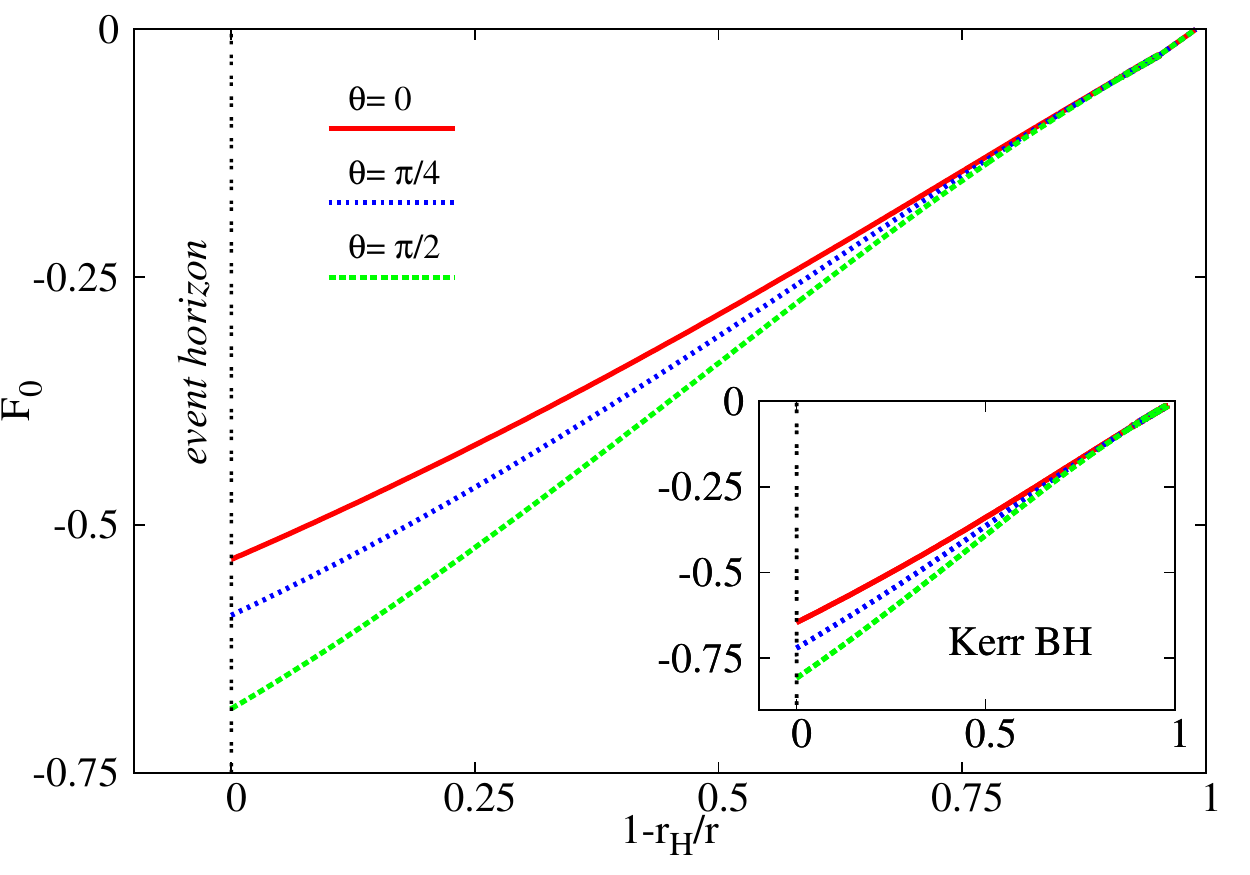} 
\includegraphics[height=.255\textheight, angle =0]{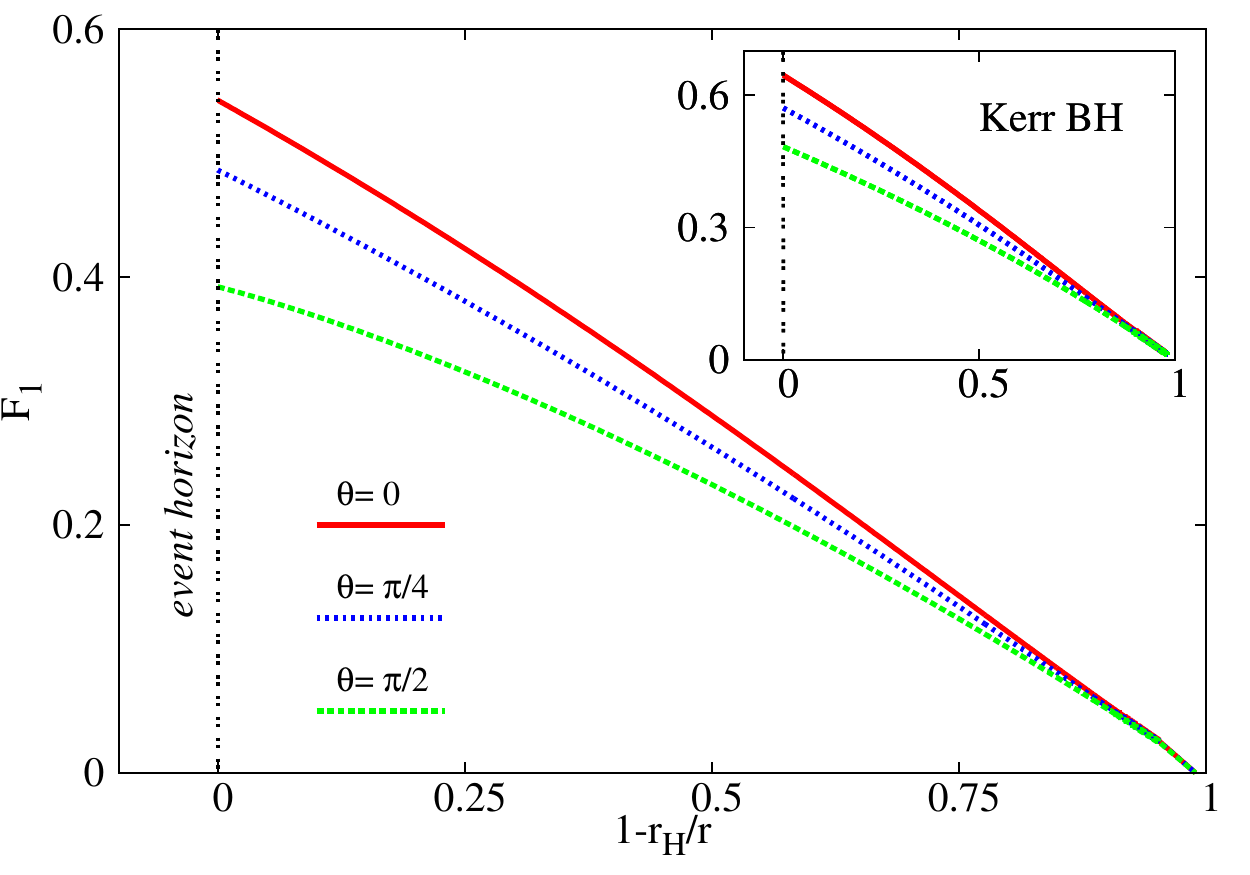} \ \ 
\includegraphics[height=.255\textheight, angle =0]{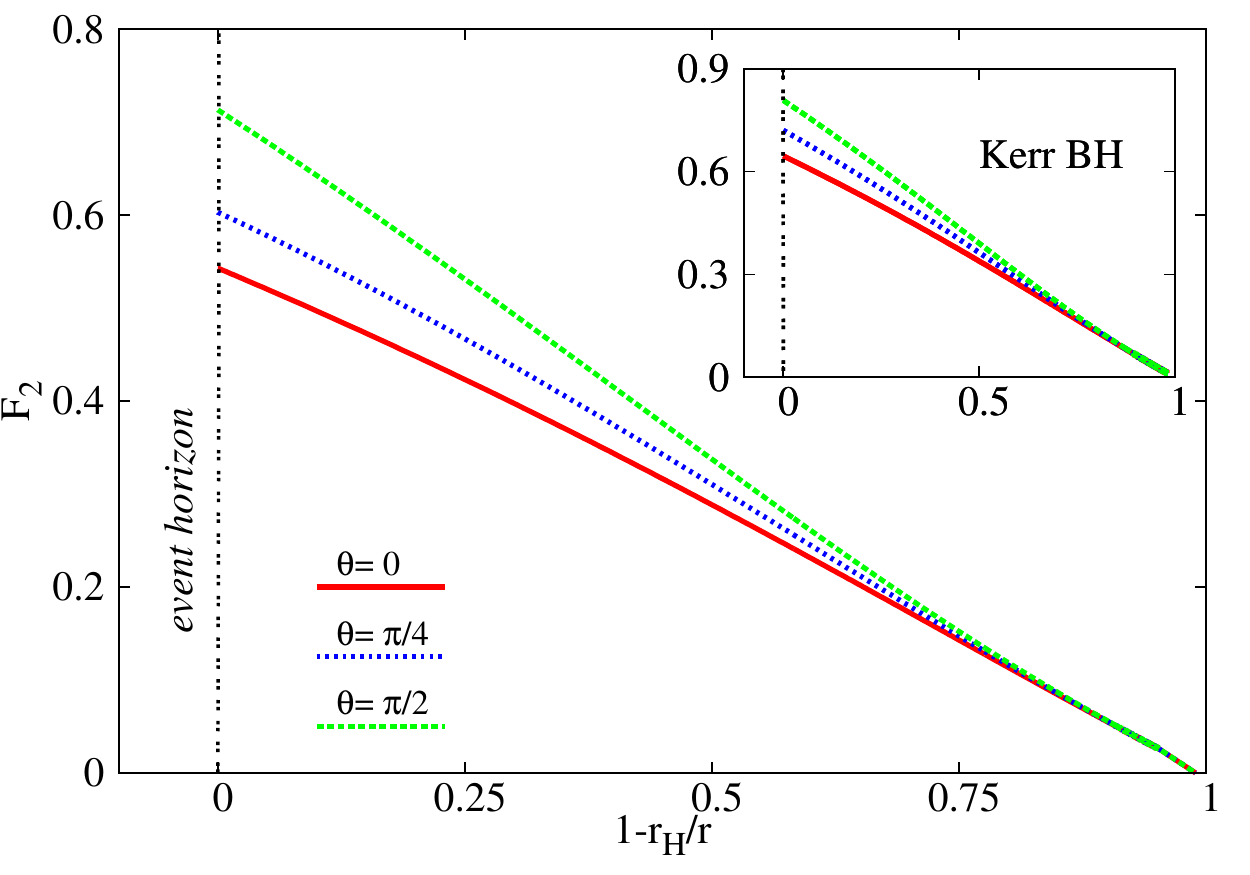} 
\includegraphics[height=.255\textheight, angle =0]{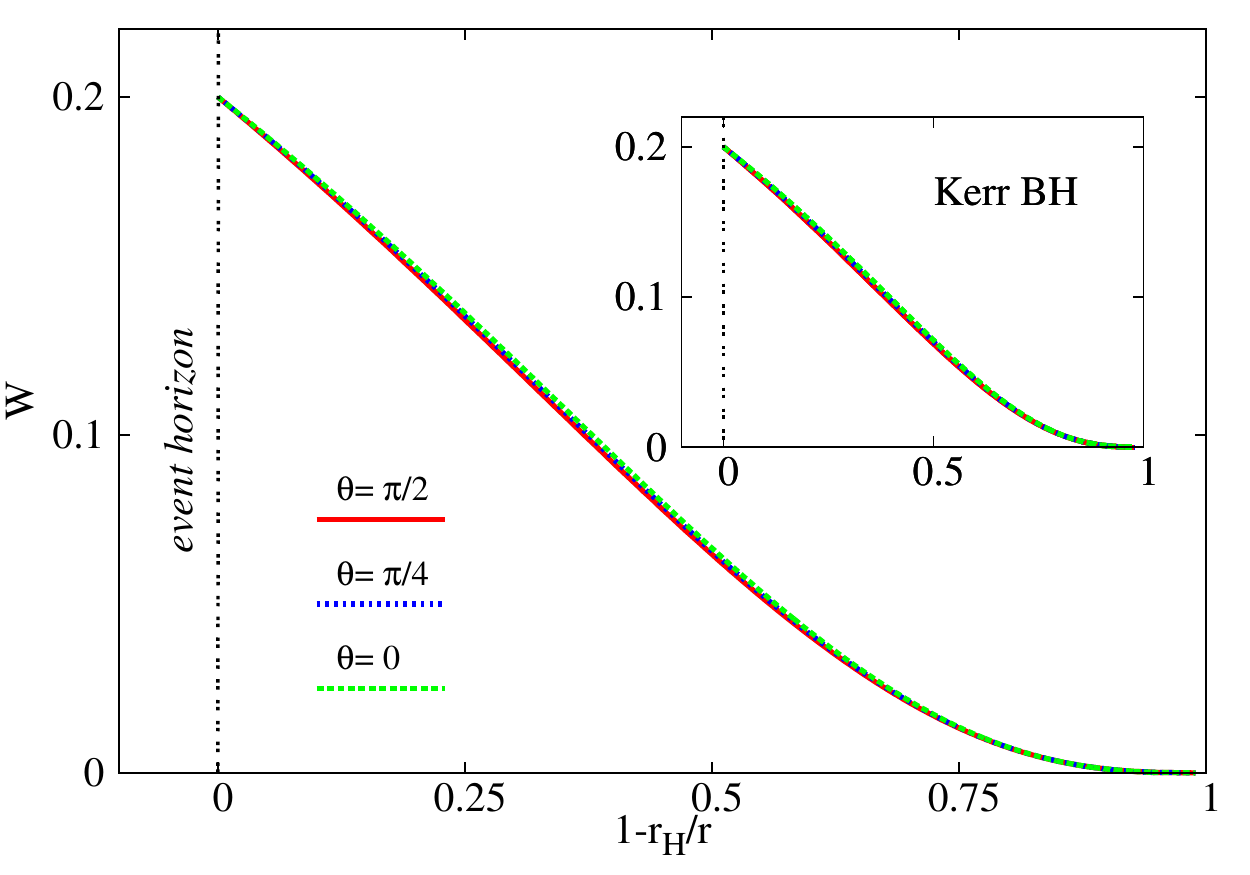} \ \
\includegraphics[height=.255\textheight, angle =0]{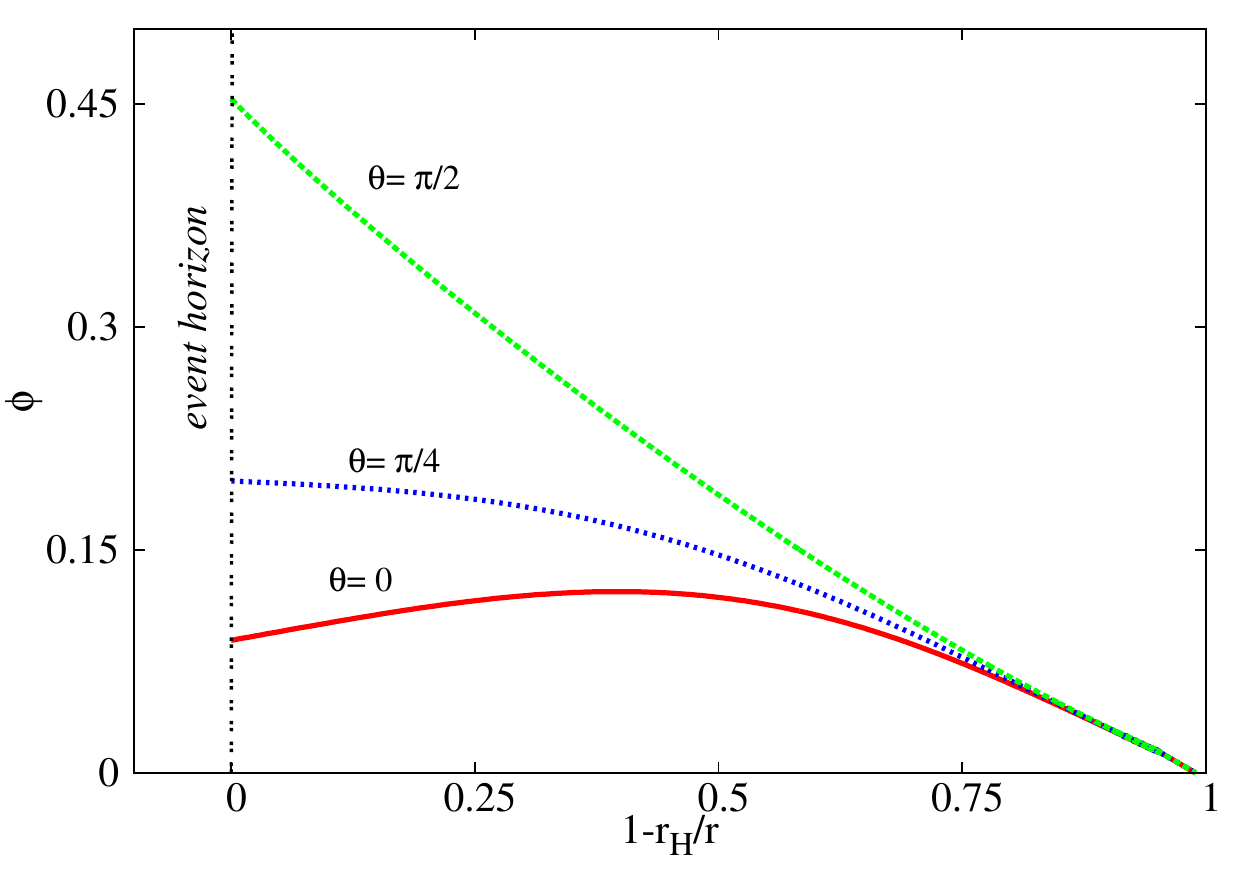} 
\includegraphics[height=.255\textheight, angle =0]{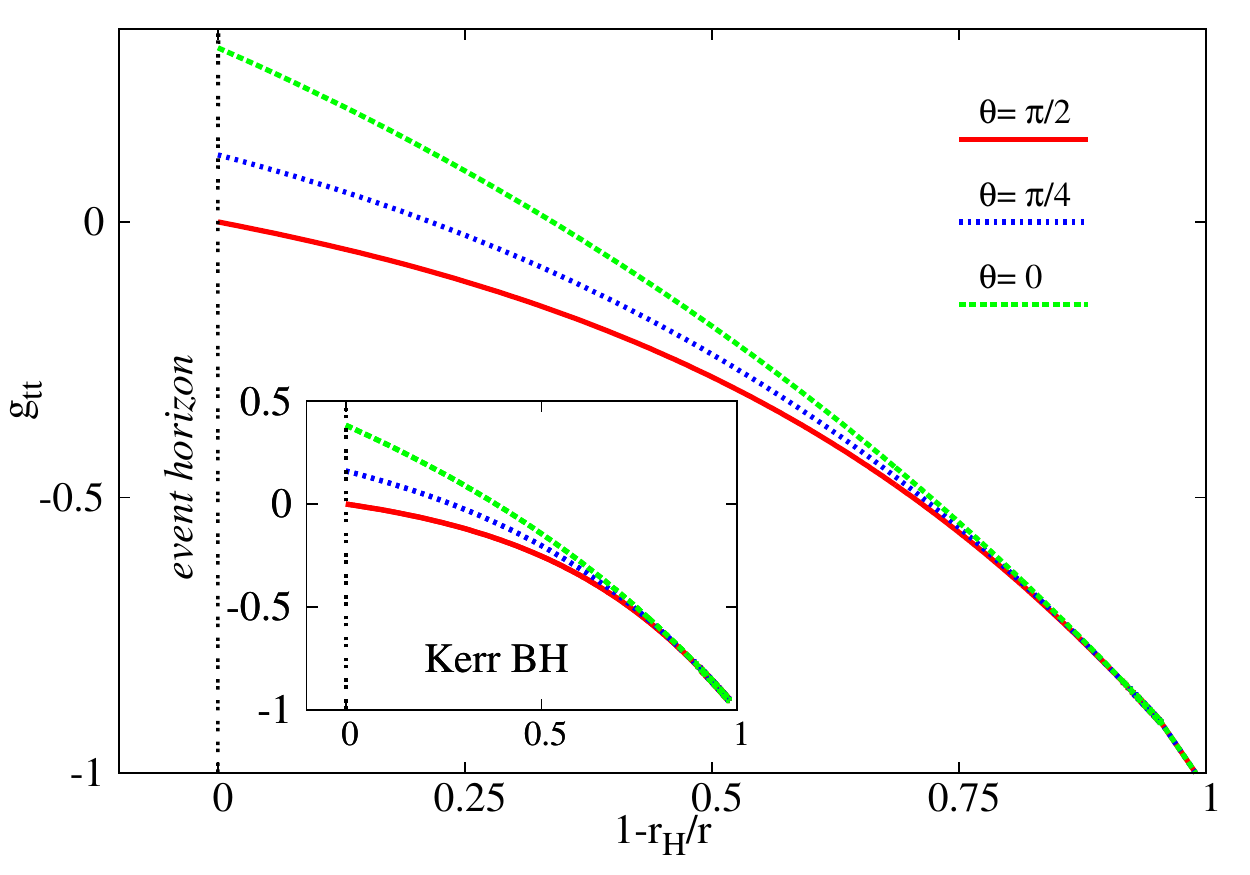} \ \  
\end{center}
  \vspace{-0.5cm}
\caption{Profile functions of a typical solution 
with 
$r_H=1.38$,
$\Omega_H=0.2$,
$\alpha=0.4$,
$vs.$ $1-r_H/r$, which compactifies the exterior region,
 for three different polar angles $\theta$.  The insets show the corresponding functions for a Kerr BH with the same $r_H,\Omega_H$. The behaviour is qualitatively similar for both cases, with small quantitative differences.
}
\label{sol1}
\end{figure}

\begin{figure}[t!]
\begin{center}
\includegraphics[height=.255\textheight, angle =0]{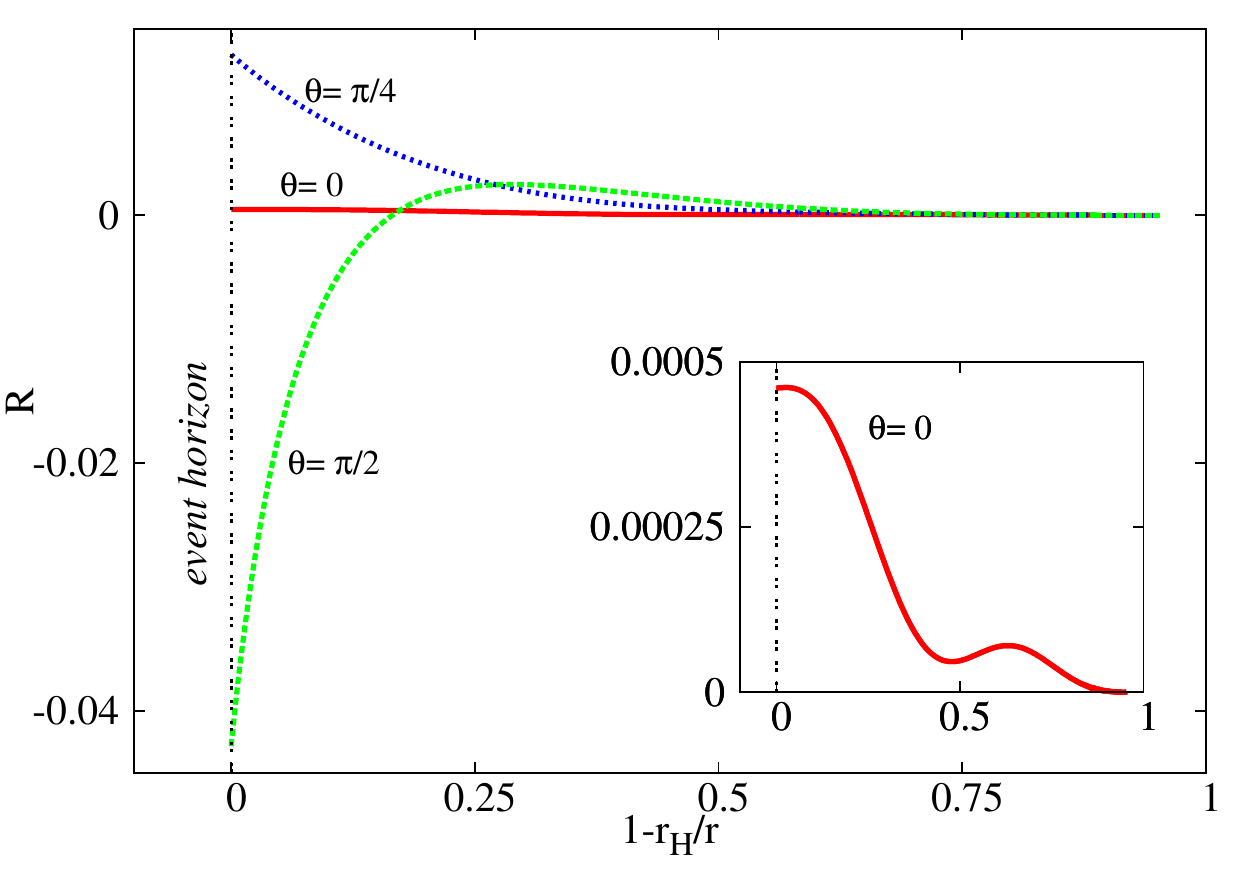} 
\includegraphics[height=.255\textheight, angle =0]{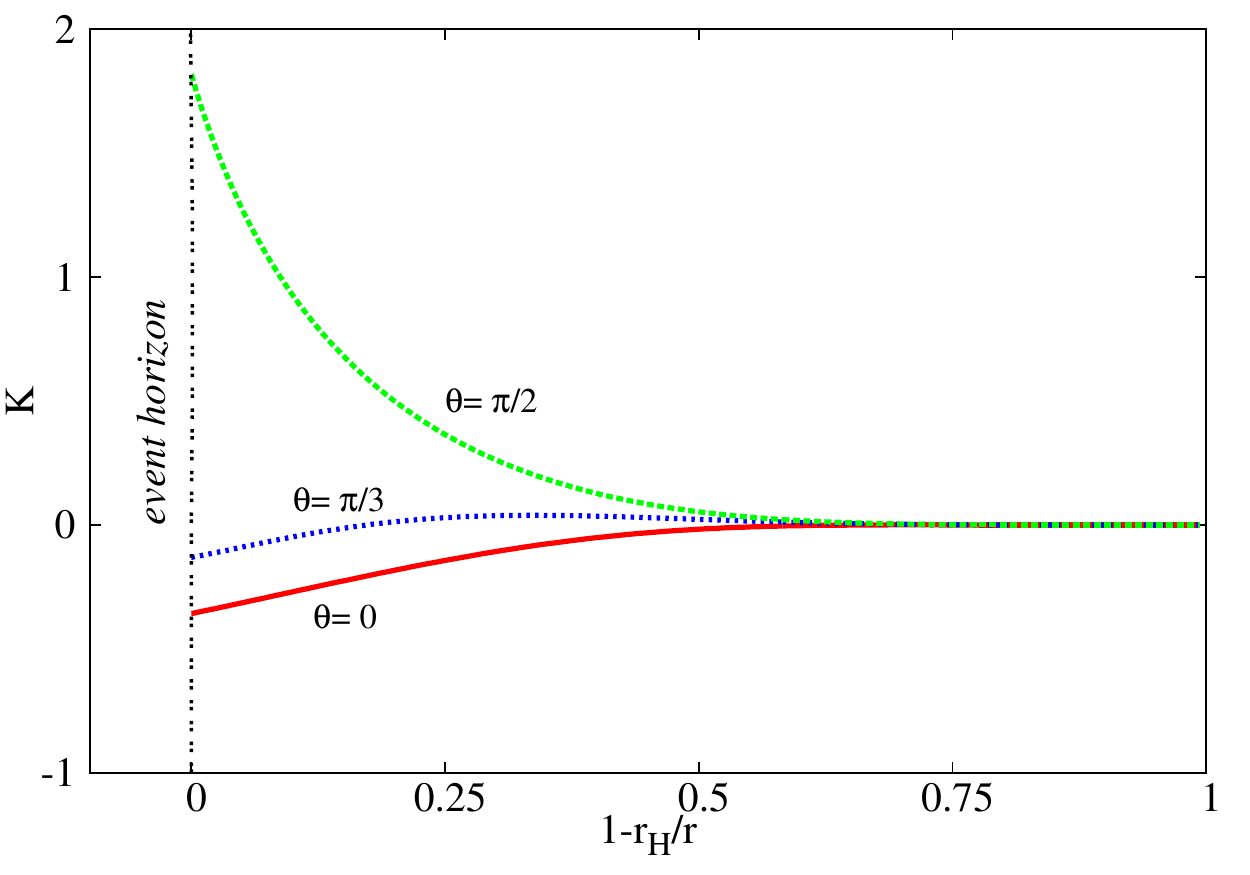}  
 \ \ 
\includegraphics[height=.255\textheight, angle =0]{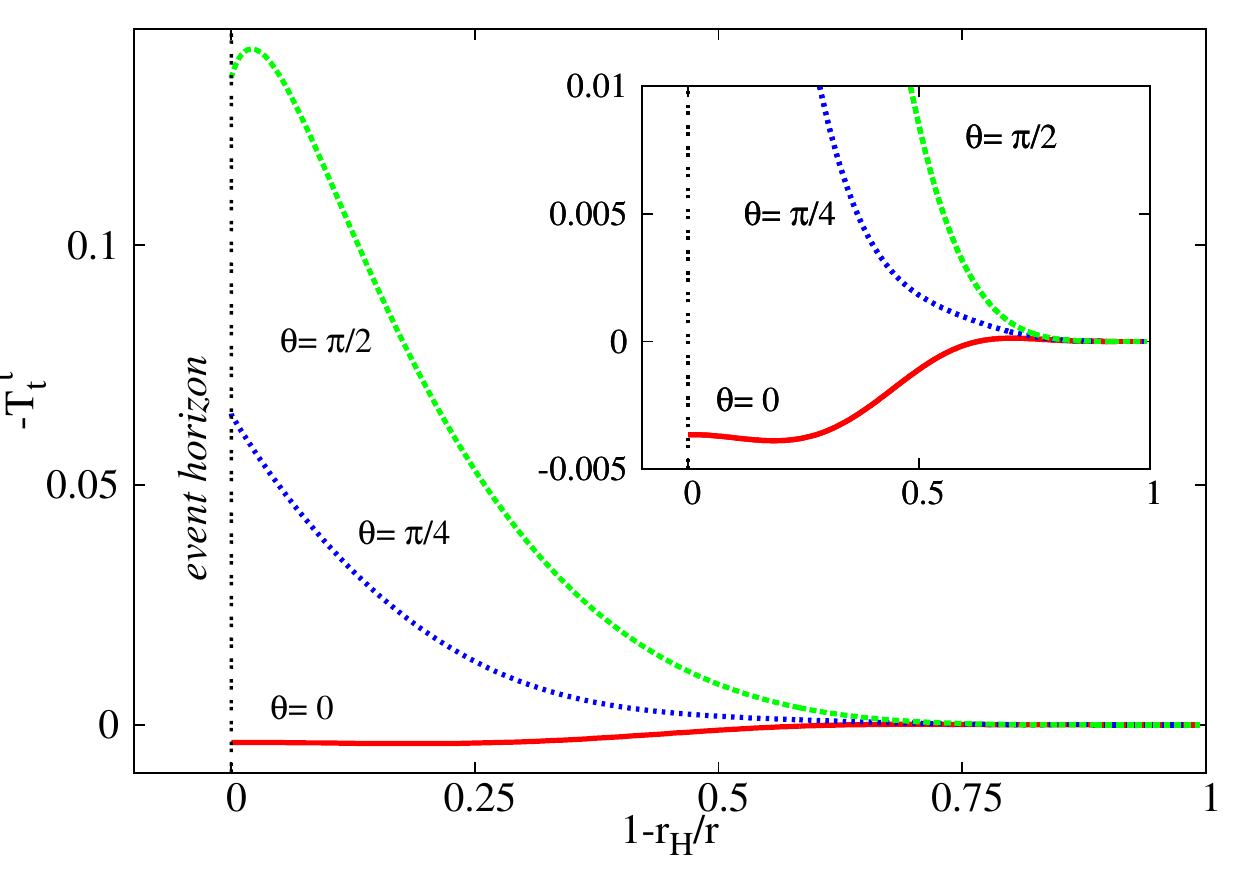} 
\includegraphics[height=.255\textheight, angle =0]{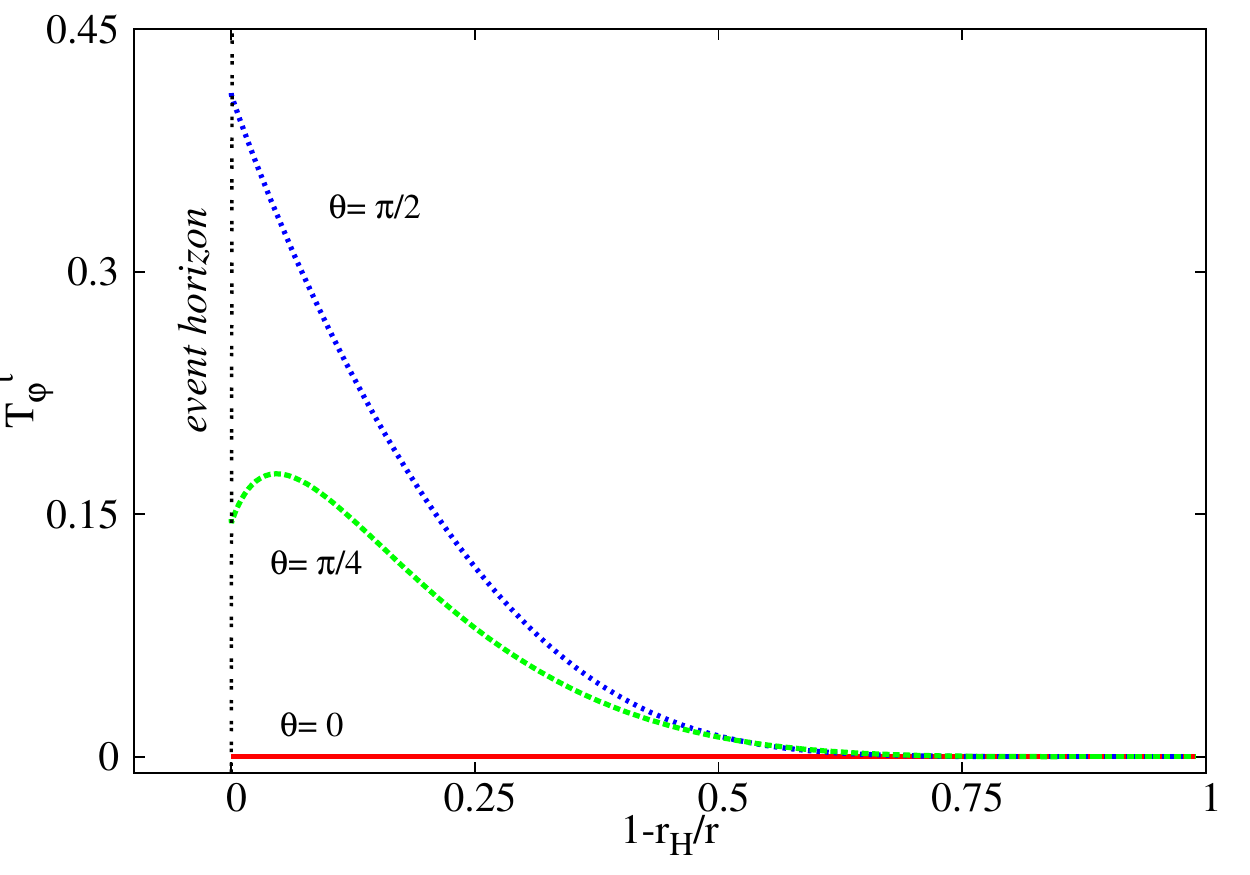} 
\end{center}
  \vspace{-0.5cm}
\caption{ The Ricci $R$ and Kretschmann  $K$  scalars
and the components $T_t^t$
and 
$T_\varphi^t$
of the {\it effective}
energy-momentum tensor, $vs.$  $1-r_H/r$, 
 for three different polar angles $\theta$ and the same solution as in Figure  
 \ref{sol1}.
The inset of the bottom left panel shows the existence of a region of negative energy densities around the axis. The inset of the top left panel shows a zoom of the $\theta=0$ curve.
}
\label{sol2}
\end{figure}

The profile functions of a typical solution are exhibited in Figure~\ref{sol1}. 
The insets show  the same curves for Kerr with the same $r_H$, $\Omega_H$, for comparison. 
The  Ricci and the Kretschmann scalars, $R$ and $K\equiv R_{\alpha\beta\mu\nu}R^{\alpha\beta\mu\nu}$, together with the components $T_t^t$ and $T_\varphi^t$
of the {\it effective} energy-momentum tensor are shown in 
 Figure \ref{sol2}.
In these plots,  the corresponding functions are shown in terms of the (inverse) radial variable
$r$
 for three different values of the
angular coordinate $\theta$.
One observes, for instance, that $g_{tt}$ becomes positive along the equator, near the horizon, thus manifesting the existence of an ergo-region (see next subsection). One also notices that both $R$ and $K$
stay finite everywhere, in particular at the horizon.
From the components of the effective energy-momentum tensor one observes, in particular,  that $-T_t^t<0$ for a region in the vicinity of the symmetry axis, manifesting a breakdown of the weak energy condition for the effective energy-momentum tensor.

Returning to the  construction of the solutions,
we have noticed the existence 
of a critical set of input parameters for which
the numerical process fails to converge.
Neither a singular behaviour nor a
  deterioration of the numerical accuracy in the vicinity of this set was observed.
 An explanation for this behaviour,  
similar to that justifying the critical configurations found in the static case,  is  based on the analysis of
the field equations in the vicinity of the event horizon.
After some algebra, one finds that the second order term $\phi_2(\theta)$ in the expansion of the 
scalar field $\phi(x,\theta)=\phi_0(\theta)+\phi_2(\theta) x^2+\dots$
is  a solution of a quadratic equation,
\begin{eqnarray}
\label{eq1}
a \phi_2^2+b \phi_2+c=0\ ,
\end{eqnarray}
where the coefficients $a,b,c$ depend on the values of $F_i,W$ and their derivatives
at the horizon.
Then, a real solution to the above equation exists only if $\Delta=b^2-4 ac>0$. 
In practice, we have monitored 
this discriminant
and
observed that  the  numerical process fails to converge\footnote{The values of $a,b,c$ becomes very large 
as the value of the reduced temperature decreases, which complicates
their accurate extraction and the evaluation of $\Delta$ in the vicinity of the extremal set.}  
 when $\Delta$  takes small values close to zero  at $ \theta=0,\pi$.
As in the spherically symmetric case, we have found no evidence for the emergence
of a secondary branch of solutions in the vicinity
of the critical solutions.

A different limiting behaviour  is found
when varying  the value of the horizon
velocity $\Omega_H$ for fixed $(r_H,\alpha)$.
As for the  vacuum Kerr family, 
following this method 
one finds two branches of solutions,
which join  for a maximal value of $\Omega_H$.
The first branch emerges from the corresponding static configuration.
The second branch, on the other hand, ends, as for $\alpha=0$,
at {\it extremal configurations}. These have vanishing Hawking temperature
and nonvanishing global charges, horizon area and entropy. We must emphasise, however, that only near extremal solutions, as opposed ot exactly extremal BHs, can be constructed within the
framework proposed in this work.
As such, the results for the extremal solutions reported here
result from extrapolating the data found in  the near-extremal case.
Moreover, unlike the extremal vacuum Kerr BH which yields a perfectly regular geometry~\cite{Bardeen:1999px},
the extremal EsGB solutions appear to not be regular, with 
the Ricci scalar tending to diverge at the poles of the horizon. 
A partial understanding of this behaviour is 
given in Appendix~\ref{apb}, based on a perturbative construction of 
the near-horizon configurations.

\subsection{The domain of existence}

\begin{figure}[ht!]
\begin{center}  
\includegraphics[height=.28\textheight, angle =0]{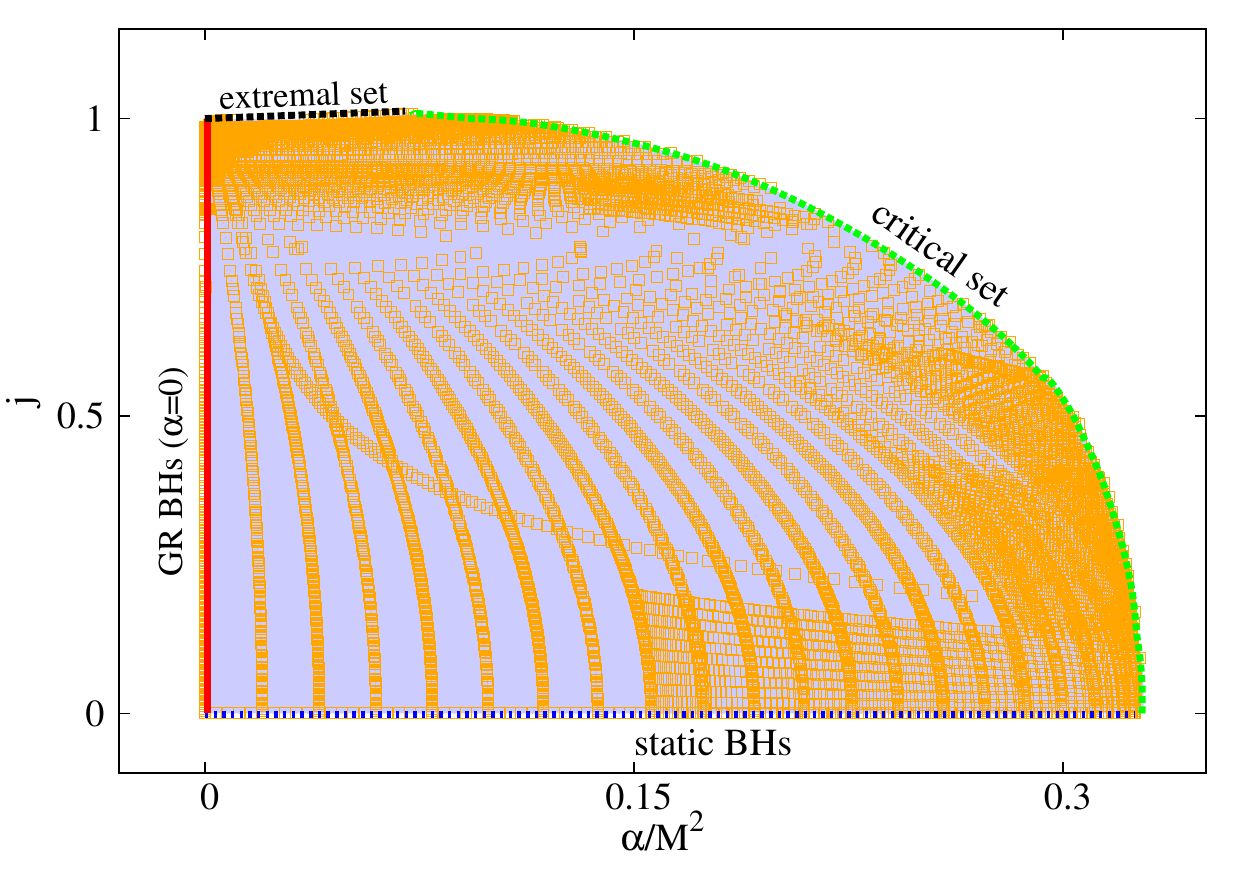} 
\end{center}
  \vspace{-0.5cm}
\caption{ Domain of existence of EsGB spinning BHs in a $j$
$vs.$~$\alpha/M^2$ diagram.
Here and in Figure 
\ref{dom1},
all quantities are normalised $w.r.t.$ the mass of the solutions. 
The domain is obtained by extrapolating into the continuum
over twenty thousand numerical points.
Each such point corresponds to an individual BH solution, and is represented in this plot as a small orange circle. 
}
\label{dom0}
\end{figure}

Let us now address the domain of existence of the EsGB solutions.
There are two fundamental scales, the coupling constant $\alpha$, 
and the BH mass of the solutions $M$.
In what follows we display various quantities of interest as a function of 
 the dimensionless coupling constant $\alpha/M^2$. This parameter measures the impact of non-GR features, due to the GB contribution. The analysis is also performed in terms of the dimensionless angular momentum $j=J/M^2$. This parameter measures the impact of non-staticity.  
The link between these two quantities is provided by the Figure \ref{dom0}, where we 
plot the domain of existence 
(shaded blue region) 
in a  $j$ $vs.$  $\alpha/M^2$ plot. Therein, all data points which were found numerically are also explicitly shown. The blue shaded region is the extrapolation of these points into the continuum. The figure shows that the domain of existence is delimited by:
 \begin{itemize}
\item  the set of static BHs ($j=0$, blue dotted line); 
\item  the set of extremal  BHs    (black dotted line);
\item  the set of critical solutions  (green line);  
\item  the set of GR solutions -- the Kerr/Schwarzschild  BHs  ($\alpha/M^2=0$, red line).
\end{itemize}

Two comments on Figure \ref{dom0}. First, the Kerr bound  $j\leqslant 1$ 
is violated for spinning EsGB BHs
in a small region of the domain of existence close the extremal set.
However, this violation
is rather small, with 
$j^{(max)}\sim 1.013$ for all (accurate enough) solutions studied so far.
Second, along $j$ fixed lines, the critical solution is attained at a smaller $\alpha/M^2$ as  $j$ is increased. A possible interpretation is that both the GB contribution and the spin are repulsive effects. Thus, in the presence of rotation, BHs cease to exist for a smaller GB contribution.

\begin{figure}[t!]
\begin{center}
\includegraphics[height=.255\textheight, angle =0]{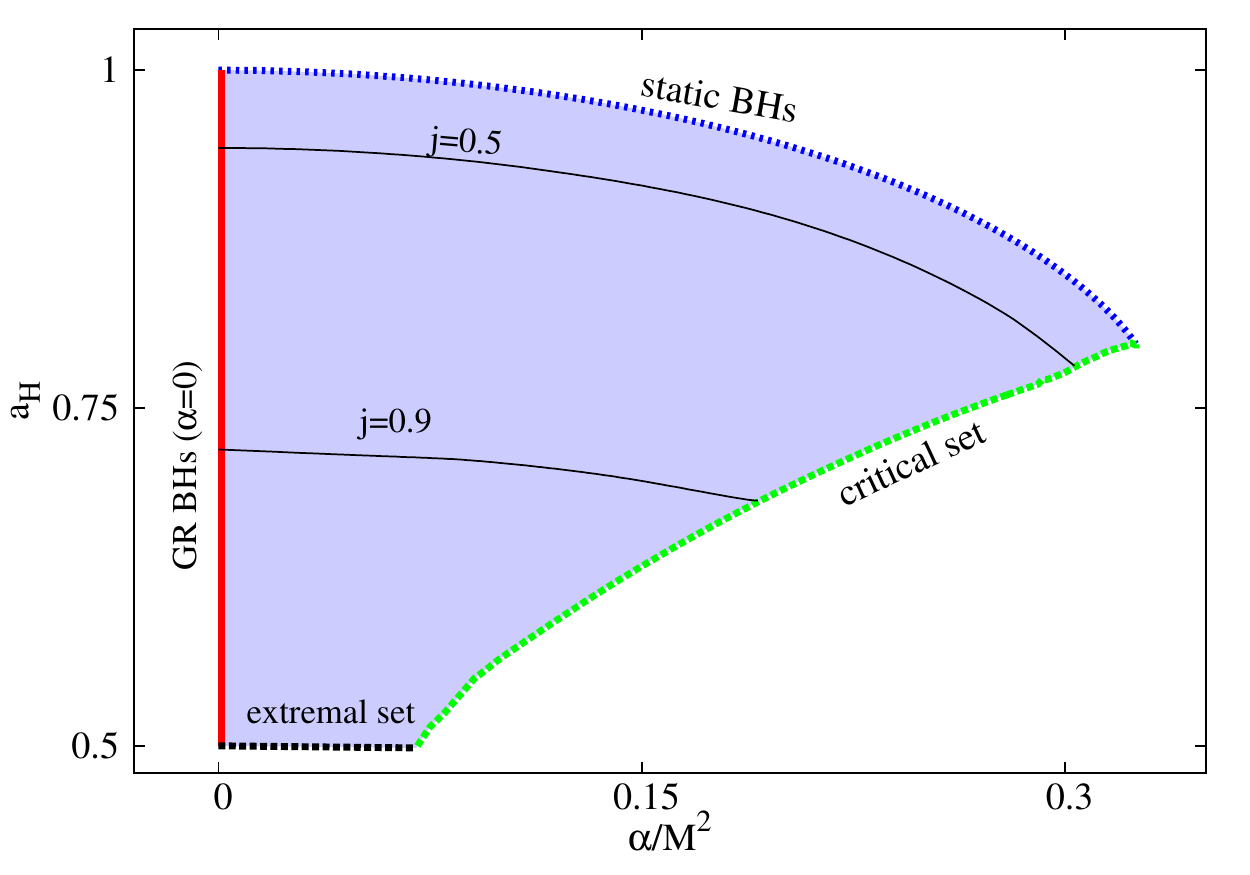} 
\includegraphics[height=.255\textheight, angle =0]{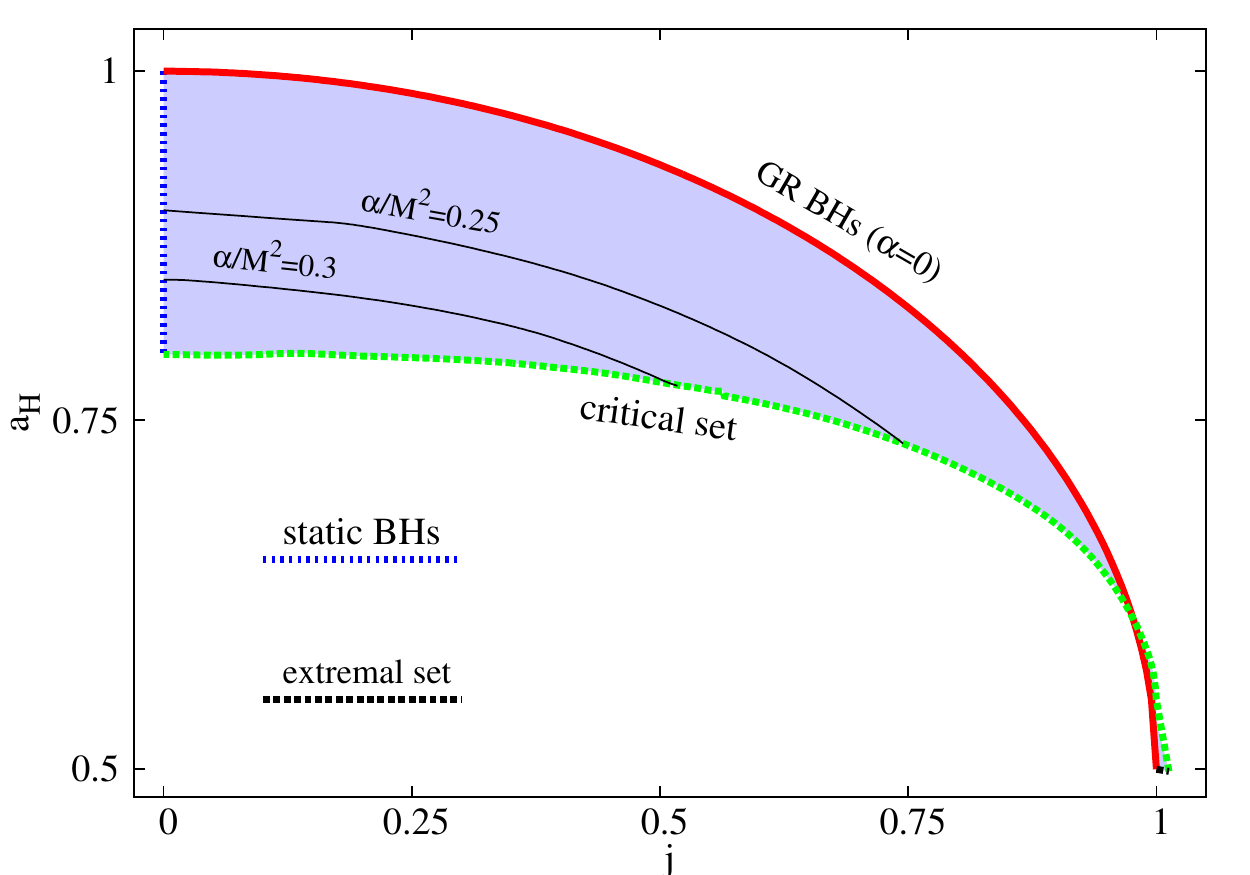} \ \ 
\includegraphics[height=.255\textheight, angle =0]{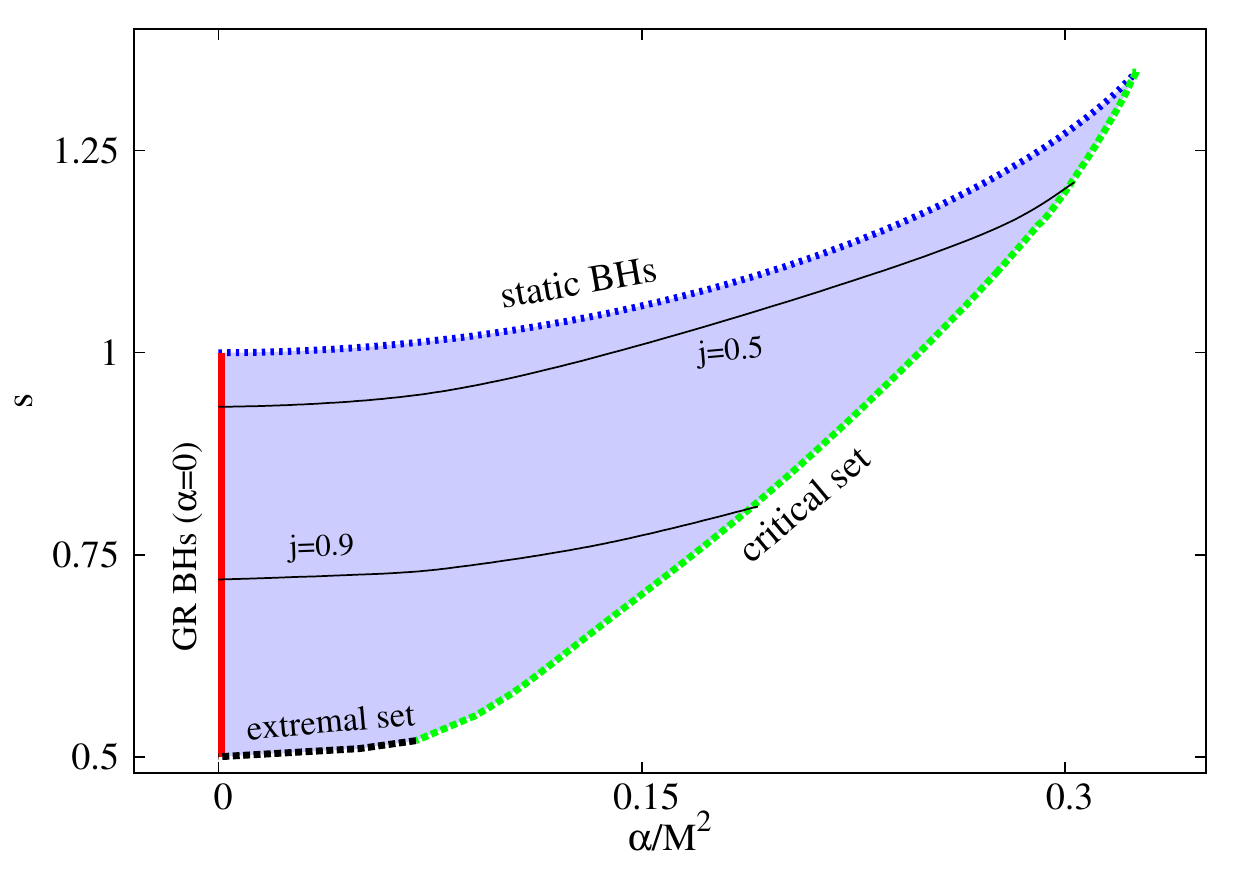} 
\includegraphics[height=.255\textheight, angle =0]{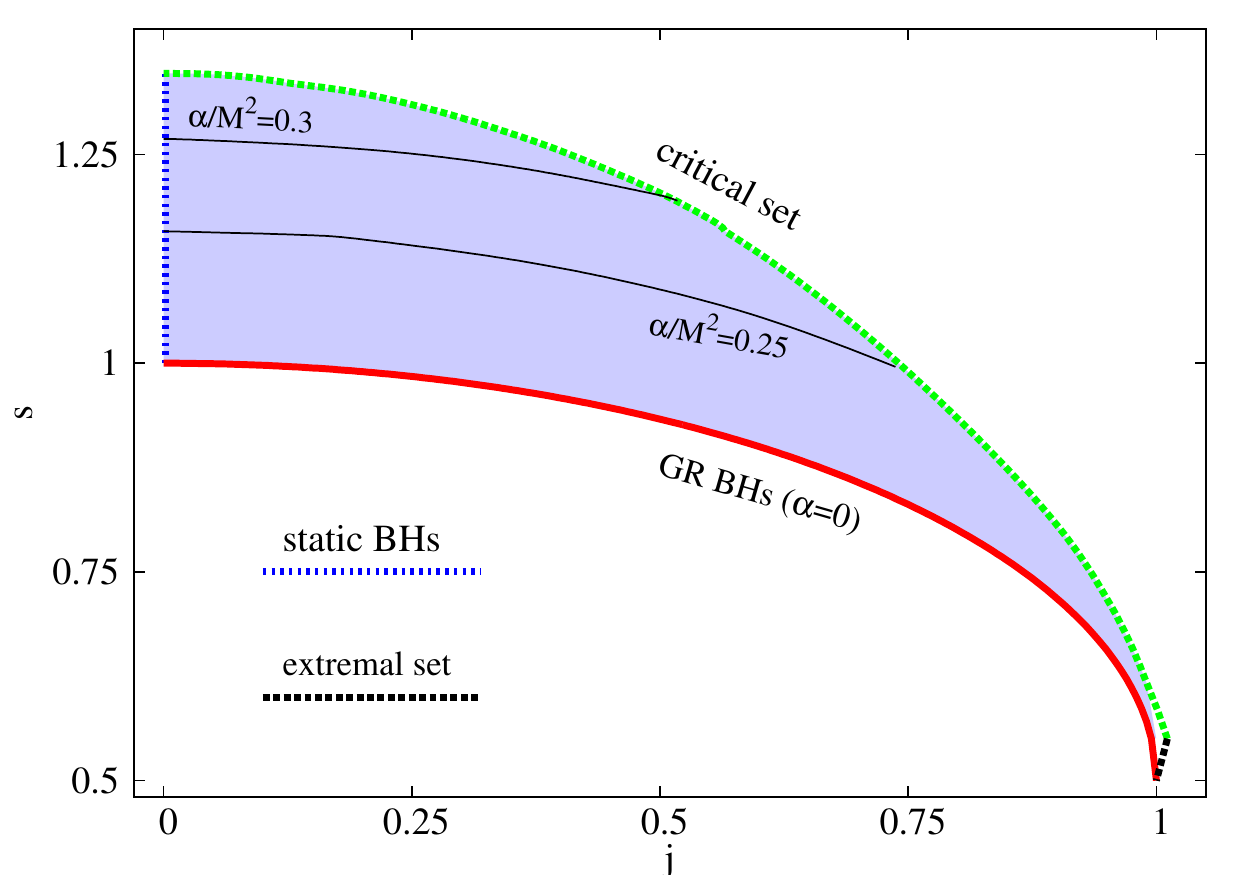} \ \
\includegraphics[height=.255\textheight, angle =0]{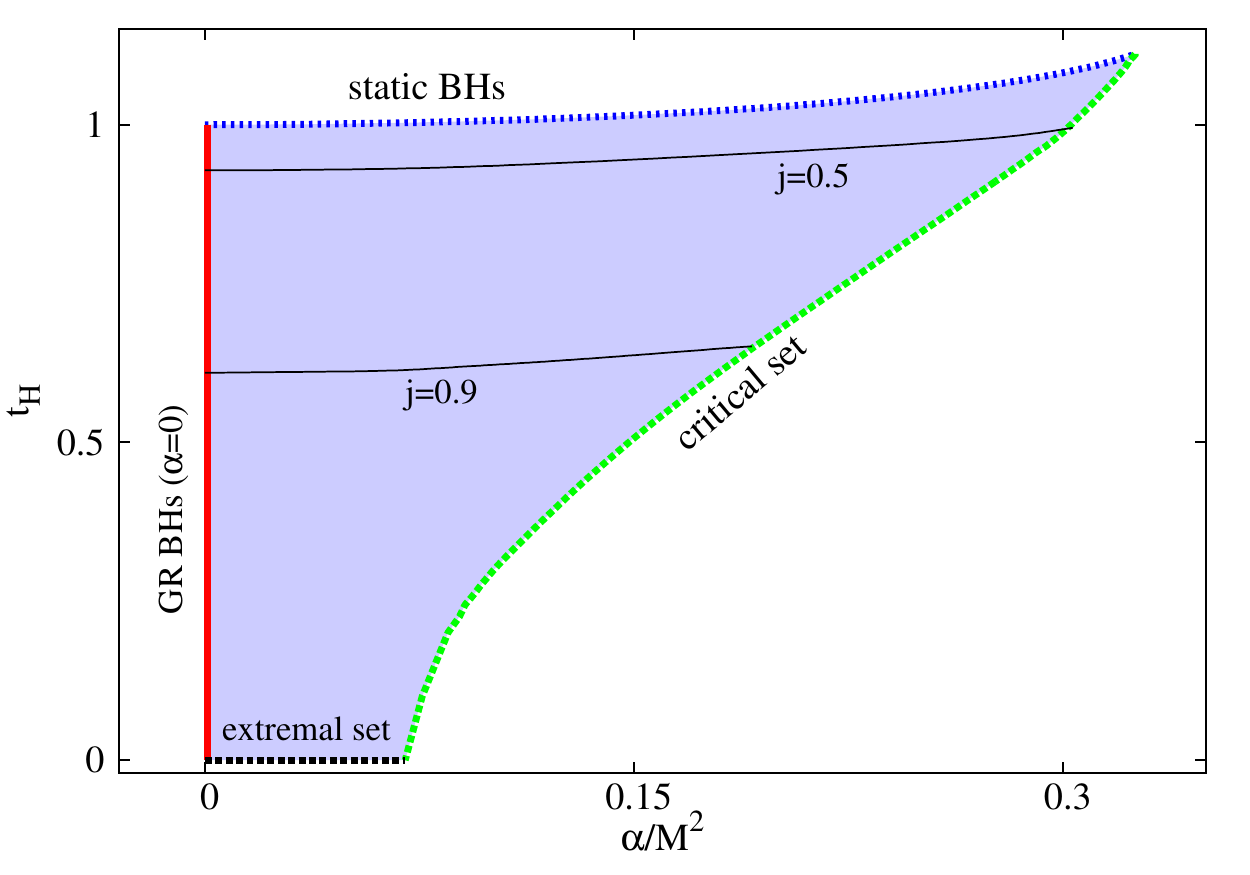} 
\includegraphics[height=.255\textheight, angle =0]{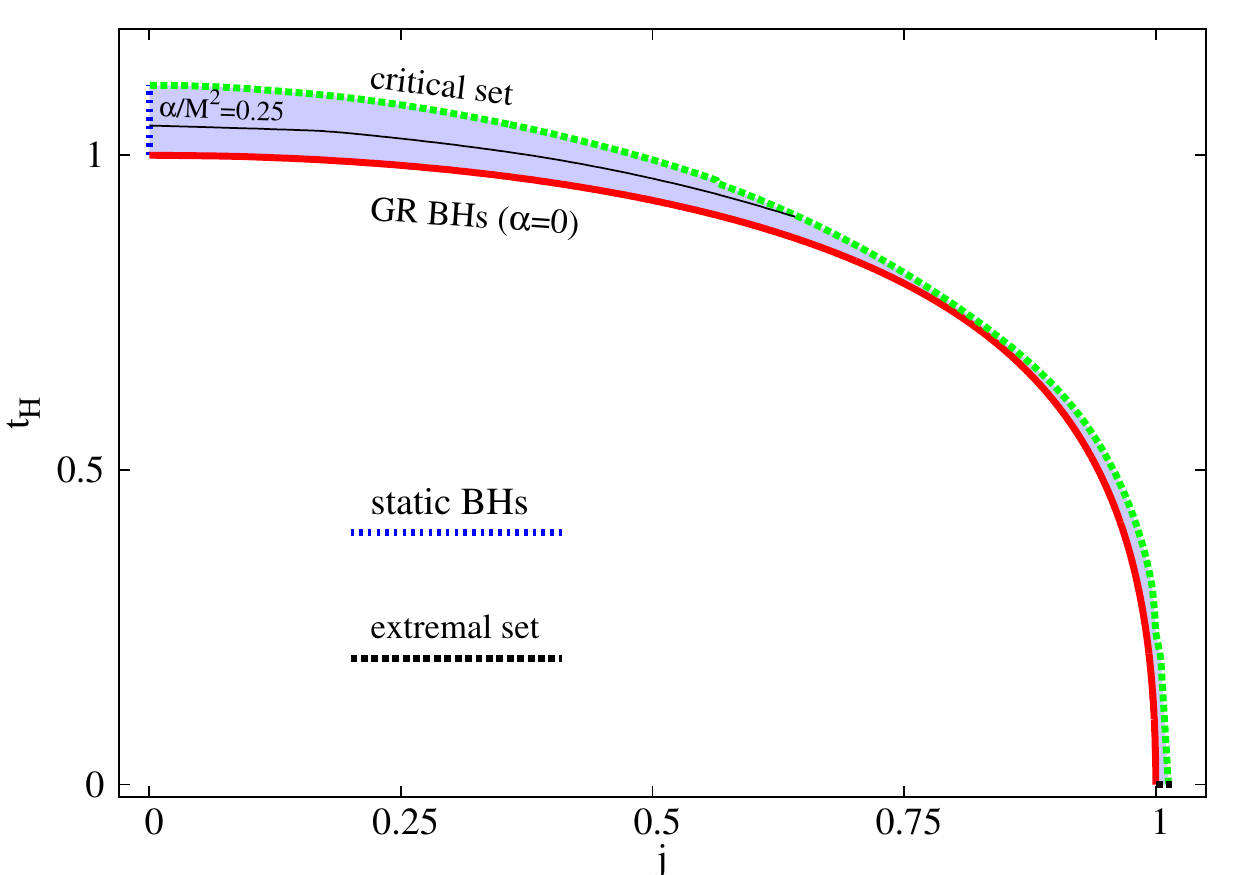} \ \  
\end{center}
  \vspace{-0.5cm}
\caption{Domain of existence of spinning EsGB BHs in a reduced horizon area (top panels), entropy (middle panels) and
Hawking temperature (bottom panels)  $vs.$ the dimensioness coupling $\alpha/M^2$ (left panels) or angular momentum $j$ (right panels).
}
\label{dom1}
\end{figure}

In Figure \ref{dom1} (left panels) 
the reduced horizon area $a_H\sim A_{\rm H}/M^2$, entropy $s\sim S/M^2$
and temperature $t_H\sim T_{\rm H} M$
of all solutions are
shown
as functions of the dimensionless coupling constant $\alpha/M^2$.
A complementary picture is found when exhibiting the same data as a function of
the reduced angular momentum $j$ -  Figure \ref{dom1}  (right panels).

Let us comment on some  features resulting from Figure \ref{dom1}. For fixed $j$, the BH area decreases as $\alpha/M^2$ increases; but the corresponding reduced BH entropy \textit{increases}. This provides a clear example how BH entropy deviates from the Hawking-Bekenstein formula in this modified gravity: when the GB contribution becomes larger, the BH becomes smaller but it carries more entropy (for fixed $j$). On the other hand, fixing the EsGB dimensionless coupling constant $\alpha/M^2$, both the reduced area and the reduced entropy decrease as $j$ increases. Thus, for any fixed EsGB model, spin reduces the size and the entropy of BHs. The BH temperature, on the other hand, increases with $\alpha/M^2$ for fixed $j$ and decreases with $j$ for fixed $\alpha/M^2$.

 \subsection{Other properties}

\subsubsection{Ergoregion and horizon properties}
All spinning EsGB BHs  have an ergoregion, defined as the domain in which the norm of $\xi=\partial_t$ becomes positive outside the horizon. 
This region is bounded by the event horizon and by the surface where
\begin{equation}
 g_{tt}=-e^{2F_0} N+W^2e^{2F_2}r^2 \sin^2\theta =0 \ .
\end{equation}
For the Kerr BH, this surface has a spherical topology and touches the horizon at the poles. 
As discussed in~\cite{Herdeiro:2014jaa}, 
the ergoregion can be more complicated for other models, notably for BHs with synchronised scalar hair, with the possible
existence of an additional $S^1\times S^1$ ergo-surface (ergo-torus) - see also~\cite{Kunz:2019bhm}. 
We have found that this is not the
case for EsGB BHs, where all solutions are Kerr-like in the sense they possess a single topologically $S^2$ ergosurface.

\begin{figure}[ht!]
\begin{center}
\includegraphics[height=.255\textheight, angle =0]{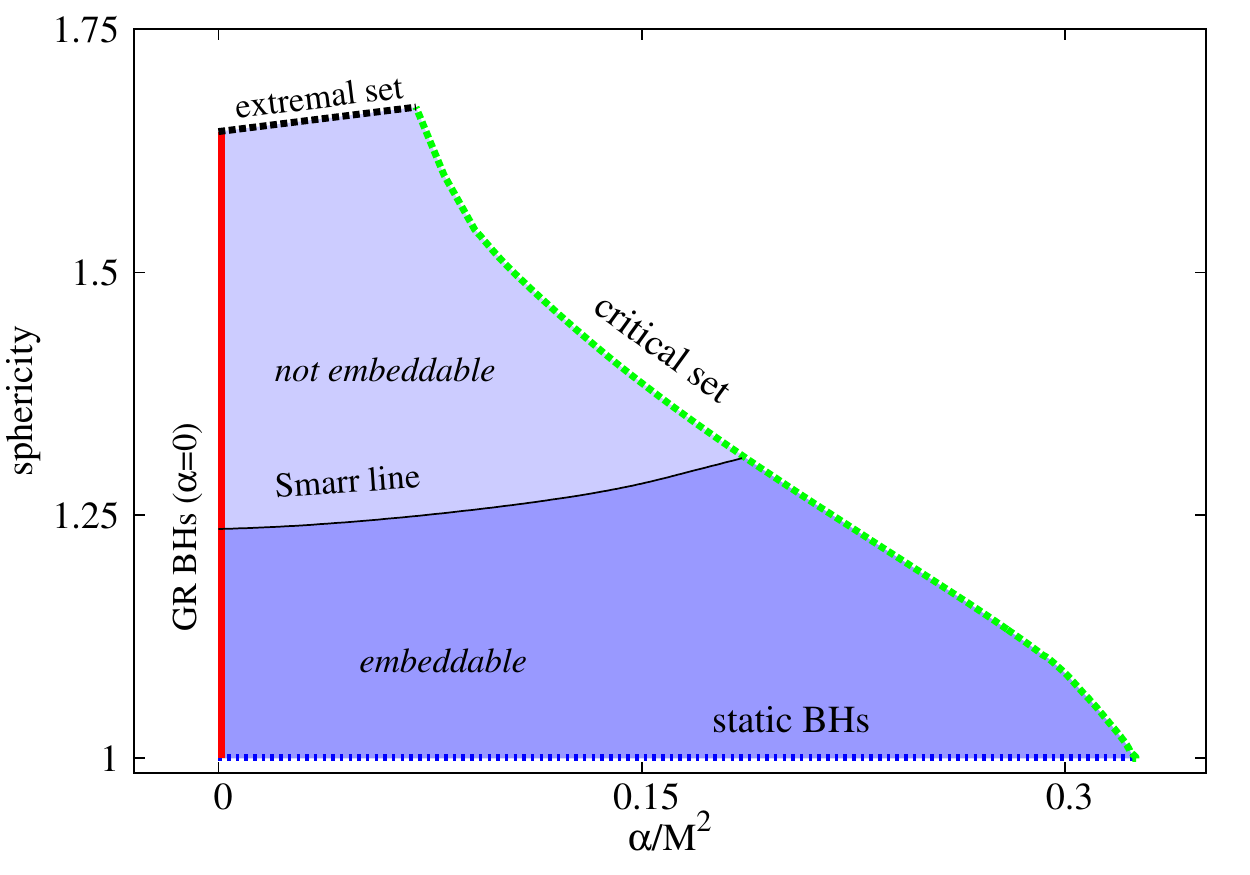} 
\includegraphics[height=.255\textheight, angle =0]{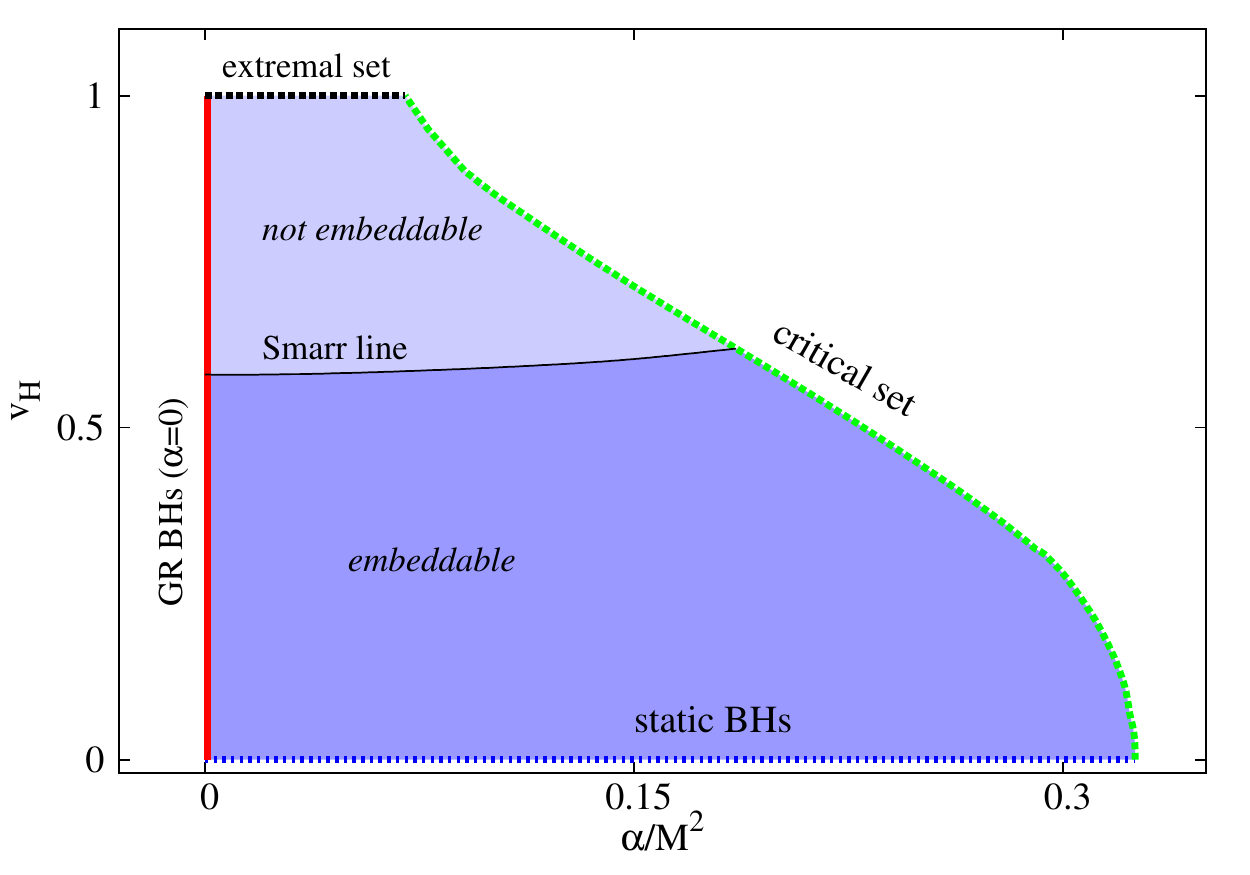}  
\end{center}
  \vspace{-0.5cm}
\caption{  The sphericity $\mathfrak{s}$ (left panel), and the horizon linear velocity $v_H$  (right panel)
$vs.$  $\alpha/M^2$ for the full set of EsGB BHs.
}
\label{Fig:Horndeski_vH}
\end{figure}

Let us now consider the horizon geometry. Similarly  to the GR Kerr solution, 
EsGB BHs have an event horizon of spherical topology. The metric of a spatial cross-section of the horizon is 
\begin{eqnarray}
\label{horizon-metric}
d\Sigma^2=h_{ij} dx^i dx^j=r_{\rm H}^2\left [ e^{2F_1(r_H,\theta)} d\theta^2+e^{2F_2(r_H,\theta)}\sin^2\theta d\varphi^2\right ]\ .
\end{eqnarray} 
Geometrically, however, the
horizon is a squashed, rather than round, sphere. 
This is shown by computing the horizon circumference  along the
equator, $L_e$, and along the poles, $L_p$:
\begin{equation}
L_e=2 \pi r_H e^{F_2(r_H,\pi/2)} \ , \qquad L_p=2 r_H \int_0^\pi d\theta e^{F_1(r_H,\theta)} \ .
\end{equation}
The ratio of these two circumferences  define the sphericity \cite{Delgado:2018khf}
\begin{equation}
\mathfrak{s} \equiv \frac{L_e}{L_p}~.
\end{equation} 
In Figure \ref{Fig:Horndeski_vH} (left panel)
the sphericity is shown as a function of the dimensionless coupling constant $\alpha/M^2$.
An interesting feature there is that $\mathfrak{s} $
can exceed the maximal GR value for a set of EsGB solutions
close to extremality. Roughly, the EsGB can become more oblate than Kerr. 
Also, as expected, the squashing of the horizon produced by the rotation is
such that $\mathfrak{s}$ is always larger than unity. That is, the solutions are always deformes towards oblatness, rather than prolatness. 

Another physical quantity of interest
 is the horizon linear velocity $v_H$~\cite{Herdeiro:2015moa,Delgado:2018khf,Delgado:2019prc}. 
 $v_H$
 measures how fast the null geodesics generators of the horizon rotate relatively to a static observer at spatial infinity.
It is defined as the product between the perimetral radius of the circumference located at the equator, 
$R_e \equiv L_e/2\pi$, and the horizon angular velocity $\Omega_H$,
\begin{equation} 
v_H=\frac{L_e }{2\pi}\Omega_H\ . 
\end{equation} 
As seen in Figure \ref{Fig:Horndeski_vH} (right panel), all studied
 EGBs solutions have  $v_H<1$, just like for Kerr, and despite the (small) violations of the Kerr bound. Thus, the null geodesics generators of the horizon rotate relatively to the asymptotic observer at subluminal speeds.

Further insight into the horizon geometry is obtained by considering the
 isometric embedding of
the spatial sections of the horizon in an Euclidean 3-space $\mathbb{E}^3$.
A well-known feature of the  Kerr horizon geometry is 
that for a dimensionless spin $j > {\sqrt{3}}/{2} \equiv j^{\rm (S)}$
(dubbed \textit{Smarr point})
the Gaussian curvature of the horizon becomes negative in a vicinity of the poles \cite{Smarr:1973zz}. 
In this regime, an isometric embedding of the Kerr horizon geometry in $\mathbb{E}^3$ is no longer possible.
As expected, this feature also occurs also for the solutions in this work,
even though the position of the Smarr point now depends on the value of 
the dimensionaless coupling constant
$\alpha/M^2$.
Following~\cite{Delgado:2018khf,Delgado:2019prc}, the collection of Smarr points as $\alpha/M^2$ is varied is dubbed {\it the Smarr line}.
Figure \ref{Fig:Horndeski_vH}  
displays also the position of the Smarr line as a function of $\alpha/M^2$.
One observes that, as for the Kerr limit, 
 an isometric embedding of the horizon geometry in $\mathbb{E}^3$ 
is possible only up to a maximal value of $\mathfrak{s} $ and $v_H$.
Also,   notice that  both the sphericity 
$\mathfrak{s}$
and
$v_H$ are not constant 
 along the Smarr line and slighly larger values of both these quantities are allowed for embeddable BHs when $\alpha/M^2$ is increased.

\subsubsection{Orbital Frequency at the ISCO and Light Rings}\label{ISCOLR}

A phenomenologically relevant aspect of any BH concerns the angular frequency at both the innermost stable circular orbit (ISCO) and the light ring (LR). 
The former is associated to a cut-off frequency of the emitted synchrotron radiation 
generated from accelerated charges in accretion disks.
 The latter is related to the real part of the frequency of BH quasi-normal modes~\cite{Cardoso:2008bp}. The LRs are also key in determining the BH shadow~\cite{Cunha:2018acu}.

Following a standard method, one finds that the angular frequency of a test particle with energy, $E$, and angular momentum, $L$, on the equatorial plane, $\theta=\pi/2$, is,
\begin{equation}
\label{Eq:AngularFrequency}
 \omega= \frac{\dot{\varphi}}{\dot{t}} = W -  \frac{e^{2(F_0-F_2)} L}{r^2 (L\ W - E)}\left( 1 - \frac{r_H}{r}\right) \ .
\end{equation}
The  radial coordinate, $r$, of such particle obeys the equation,
\begin{equation}
\dot{r}^2 = V(r) \equiv e^{-2F_1} \left( 1 - \frac{r_H}{r} \right) \left[ -\epsilon - e^{-2F_2} \frac{L^2}{r^2} + \frac{e^{-2F_0}  (E - L\ W)}{1 - \frac{r_H}{r}} \right] \ ,
\end{equation}
where the `dot' denotes derivative with respect to an affine parameter. $\epsilon$ is a constant with $\epsilon = 0$ for massless test particles and $\epsilon = -1$ for the massive test particles. The former are relevant for the LRs and the latter for the ISCO.

In the case of massive test particles, circular orbits require that both the potential $V(r)$ and its derivative vanish, $V(r) = V'(r) = 0$. This yields two algebraic equations for $E$ and $L$, which can be solved analytically. These have two distinct pairs of solutions, $(E_+, L_+)$ and $(E_-, L_-)$, corresponding, respectively, to co-rotating and counter-rotating orbits.
It is then possible to assess the stability of the circular orbits by computing the second derivative of the potential. The ISCO will correspond to the orbit in which the test particle has energy and angular momentum that solves $V(r) = V'(r) = 0$ and the radial coordinate that solves $V''(r)=0$. Having obtained the energy, angular momentum and radial coordinate of the ISCO,  the corresponding angular frequency is computed using~\eqref{Eq:AngularFrequency}.

\begin{figure}[ht!]
\begin{center}
\includegraphics[height=.255\textheight, angle =0]{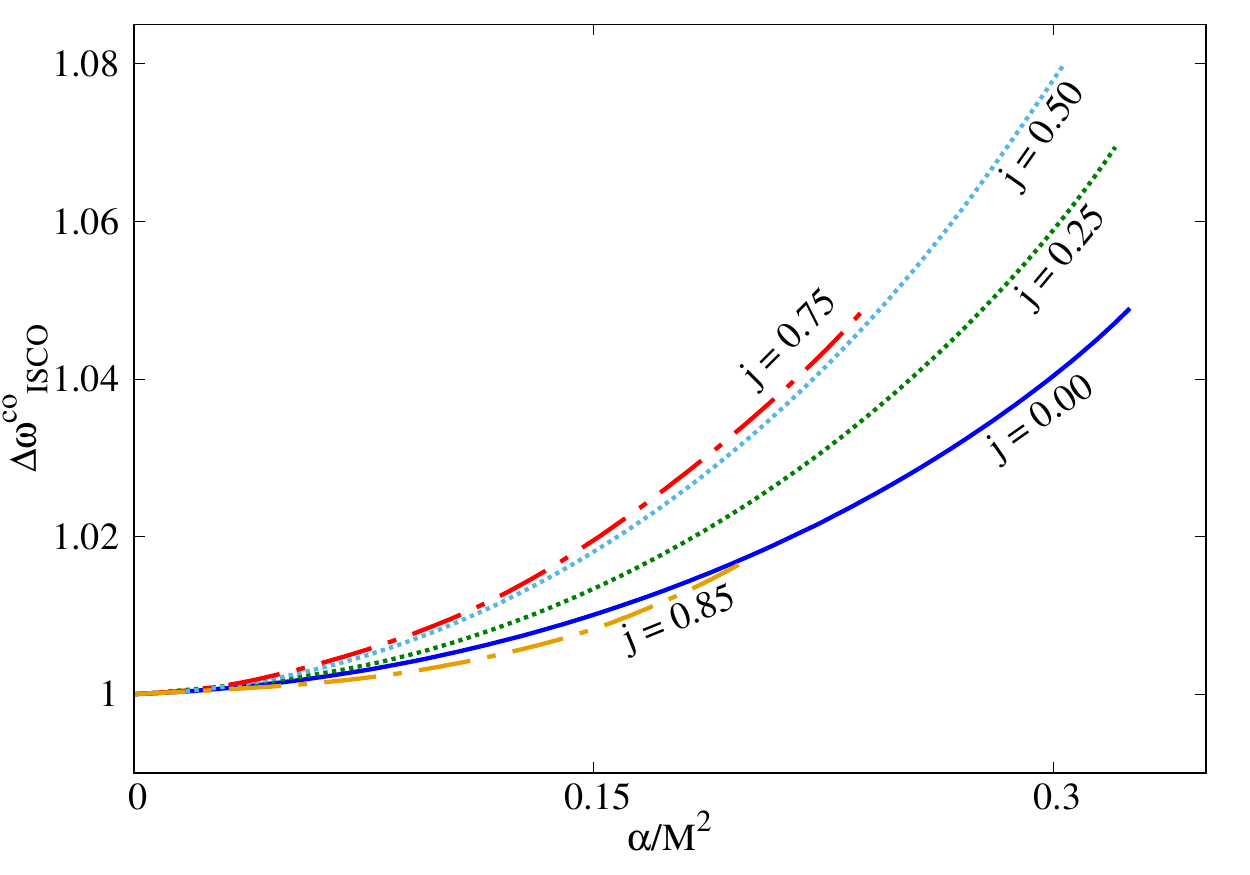} 
\includegraphics[height=.255\textheight, angle =0]{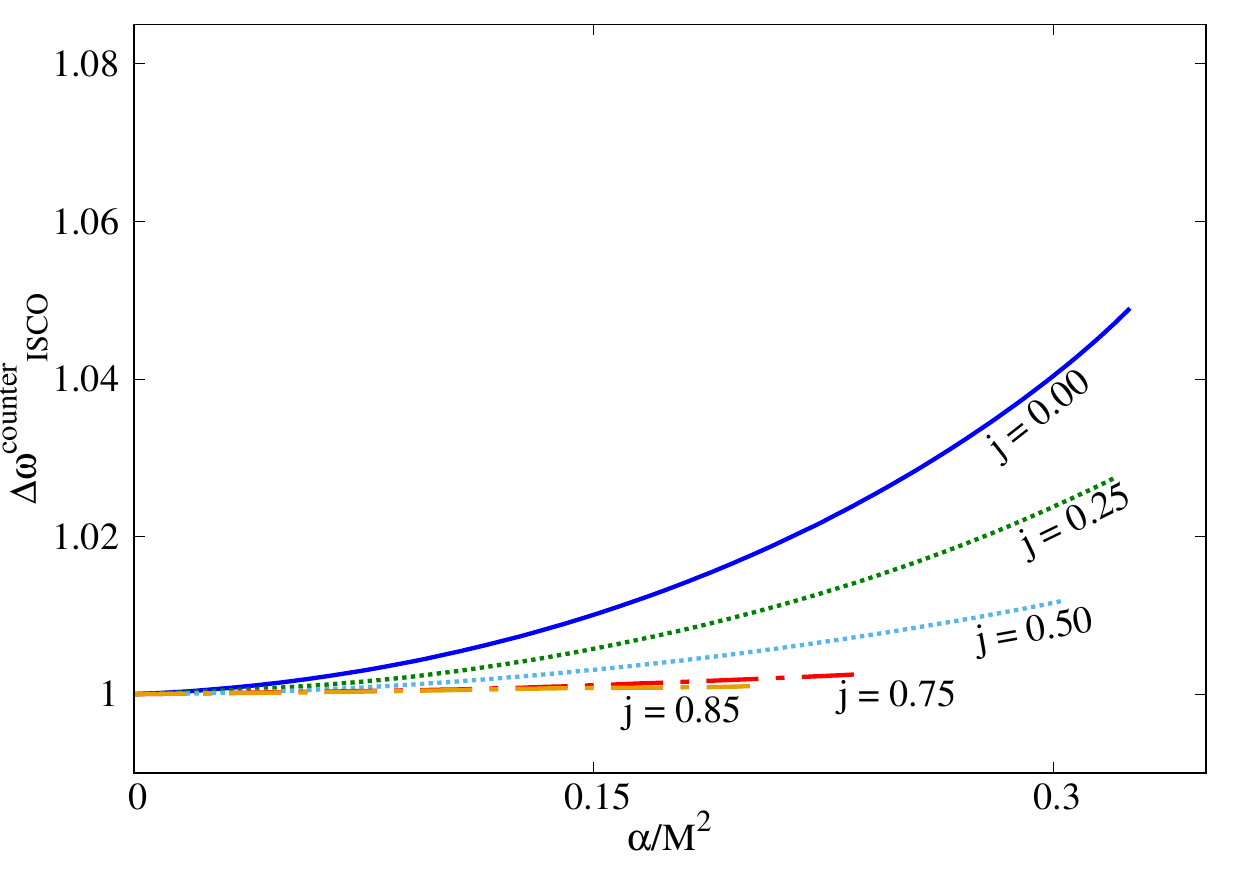}  
\end{center}
  \vspace{-0.5cm}
\caption{ Ratio between the angular frequency at the ISCO between EsGB BHs and Kerr BHs for co-rotating orbits (left panel)
and  counter-rotating orbits (right panel).}
\label{Fig:Horndeski_wISCO}
\end{figure}

In Fig.~\ref{Fig:Horndeski_wISCO}, we present the ratio between the angular frequency at the ISCO between EsGB BHs and Kerr BHs,
 for both co-rotating, $\Delta \omega_{\rm ISCO}^{\text{co}}$  and counter-rotating orbits, $\Delta \omega_{\rm ISCO}^{\text{counter}}$, 
fixing $j$, as a function of the reduced coupling constant, $\alpha/M^2$:
\begin{equation}
\label{ratios}
\Delta \omega_{\rm ISCO}^{\text{co}}(j,\alpha/M^2)=\frac{\omega_{\rm ISCO}^{\text{co}}(j,\alpha/M^2)}{\omega_{\rm ISCO}^{\text{co}}(j,\alpha/M^2=0)} \ , \qquad \Delta \omega_{\rm ISCO}^{\text{counter}}(j,\alpha/M^2)=\frac{\omega_{\rm ISCO}^{\text{counter}}(j,\alpha/M^2)}{\omega_{\rm ISCO}^{\text{counter}}(j,\alpha/M^2=0)} \ .
\end{equation}
Several illustrative values of $j$ are exhibited.

For both the co-rotating and counter-rotating cases, by definition, the ratio converges to unity in the Kerr limit. For all fixed $j$ and for both co and counter-rotating orbits, the ratio diverges away, monotonically, from unity as $\alpha/M^2$ increases. How the ratio goes away from unity depends, however, on $j$ and on the direction of the orbital motion. 

For $j=0$ the distinction between co and counter rotating orbits is meaningless. The ratio grows away from unity as  $\alpha/M^2$ increases -- solid blue line in Fig. \ref{Fig:Horndeski_wISCO}. Naively, this is related to the fact that the static BH size decreases with increasing $\alpha/M^2$, making the ISCO also decrease and hence its frequency increase. Introducing $j$ raises the degeneracy between co and counter rotating orbits. For co-rotating (counter-rotating) orbits and small $j\neq 0$, the ratio is always larger (smaller) than that for the static BHs ($j=0$) -- dotted lines in Fig. \ref{Fig:Horndeski_wISCO} (left and right panels). One may interpret these behaviours as a consequence of frame dragging, which enhances (damps) motion along co-rotating (counter-rotating) orbits. In the counter-rotating case this trend remains for large $j$ -- dashed lines in Fig. \ref{Fig:Horndeski_wISCO} (right panel). In the co-rotating case, however, an unexpected behaviour emerges.  For sufficiently large $j$ the ratio stops being enhanced with respect to the static case, and eventually becomes \textit{suppressed} with respect to it -- dashed lines in Fig. \ref{Fig:Horndeski_wISCO} (left panel).

\begin{figure}[ht!]
\begin{center}
\includegraphics[height=.255\textheight, angle =0]{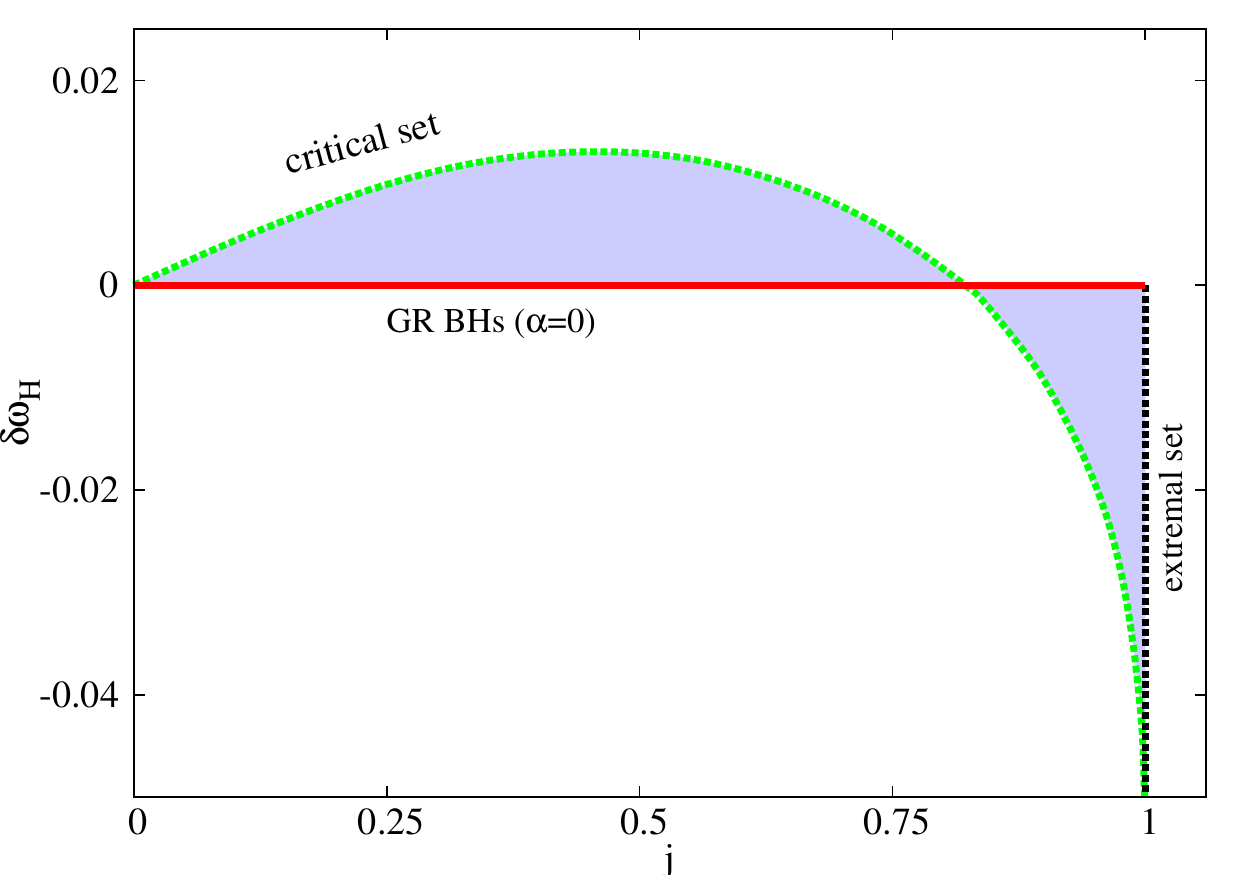} 
\end{center}
  \vspace{-0.5cm}
\caption{ Reduced horizon angular velocity difference between EsGB BHs and Kerr BHs in a $\delta \omega_H$ $vs.$ $j$ plot. For small $j$  the difference is positive, meaning that EsGB BHs spin faster. But for large $j$ the difference is negative, meaning that EsGB BHs spin slower.}
\label{Fig:Horndeski_deltaOmegaH}
\end{figure}

 A possible explanation for this unexpected behaviour is found by studying the angular velocity of the horizon, $\Omega_H$. This quantity is a better measure of dragging effects than the spacetime angular momentum. Indeed, the fact that a BH has a large $j$ does not imply that it has a large horizon angular velocity.\footnote{The relation between the two quantities should be determined by a moment of inertia. See~\cite{Herdeiro:2009qy} for an attempt to introduce this notion in BH physics.}  Let us then consider the reduced horizon angular velocity, $\omega_H \equiv \Omega_H M$, and its difference beween EsGB and Kerr BHs with the same $j$, defined as:
\begin{equation}
\delta \omega_H(j,\alpha/M^2) \equiv \omega_H(j,\alpha/M^2) - \omega_H(j,\alpha/M^2 = 0)  \ .
\end{equation} 
This quantity is plotted against the reduced angular momentum $j$ in Fig. \ref{Fig:Horndeski_deltaOmegaH}. One observes that, for small enough fixed $j$, the EsGB BHs have larger $\omega_H$ than Kerr ones. This support the thesis that dragging effects are stronger and should enhance the angular frequency at the ISCO. However, after a given spin $j$, the EsGB BHs have smaller $\omega_H$ than Kerr BHs. That is, albeit having a larger spacetime angular momentum, large $j$ EsGB BHs spin more slowly, and thus source weaker frame dragging, than Kerr BHs. Qualitatively, at least, this provides an explanation for the behaviour observed in Fig. \ref{Fig:Horndeski_wISCO} (left panel).


Quantitatively, for co-rotating orbits, the maximal deviation from Kerr is $\Delta \omega_{\rm ISCO}^{\text{co}}\sim 8\%$ and occurs for  $j \sim 0.5$ and the maximal value of $\alpha/M^2$. For counter-rotating orbits, on the other hand, the ratio is maximised, for any $\alpha/M^2$, by the static case.

In the case of massless particles, a similar analysis can be done. Now, solving $V(r) = 0$, we obtain an algebraic equation for the impact parameter, $b_p = L/E$, which yields two distinct solutions $b_p^+$ and $b_p^-$ corresponding to co-rotating and counter-rotating orbits, respectively. Using this result, and solving $V'(r) = 0$, yields the radial coordinate of the LR. Having computed the impact parameter and the radial coordinate of the LRs, one can again compute their angular frequency, using~\eqref{Eq:AngularFrequency}.

\begin{figure}[ht!]
\begin{center}
\includegraphics[height=.255\textheight, angle =0]{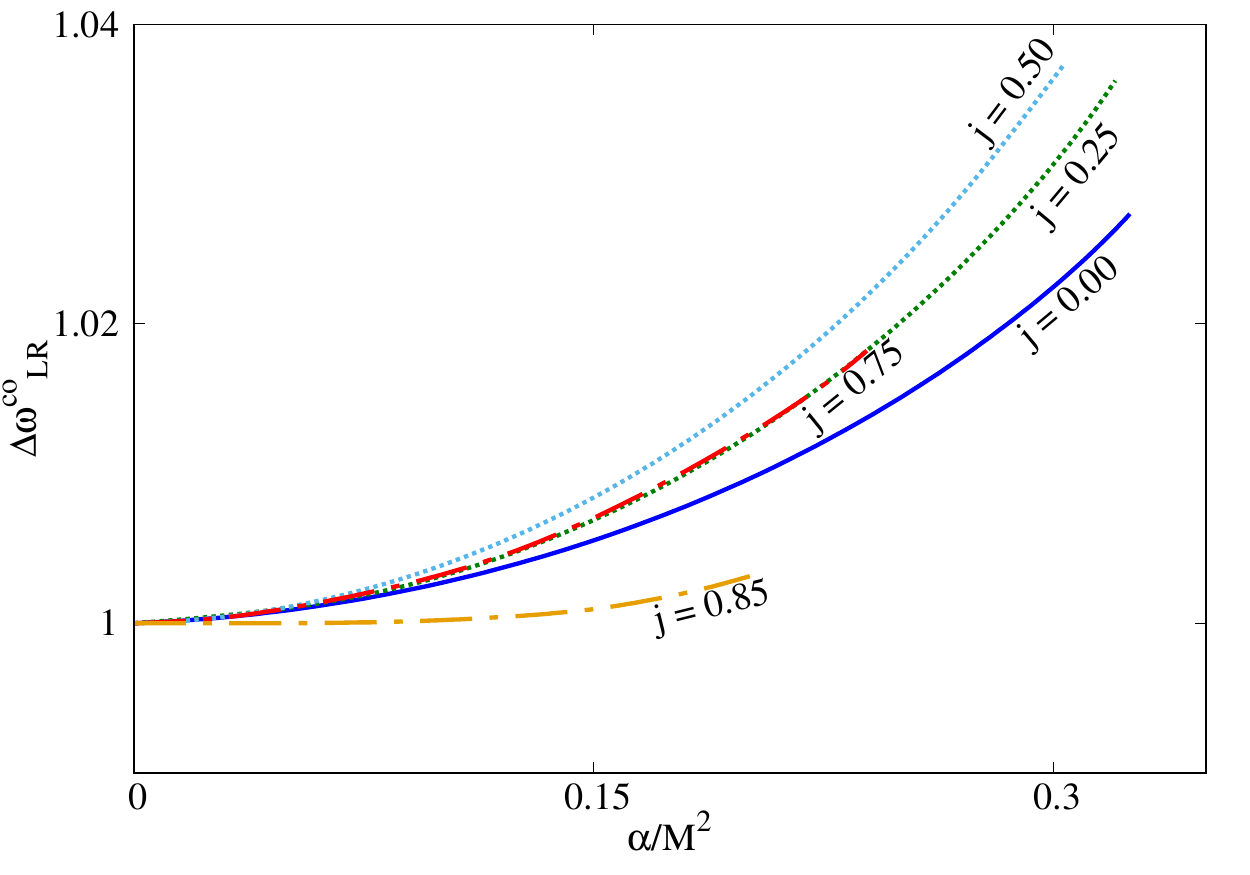} 
\includegraphics[height=.255\textheight, angle =0]{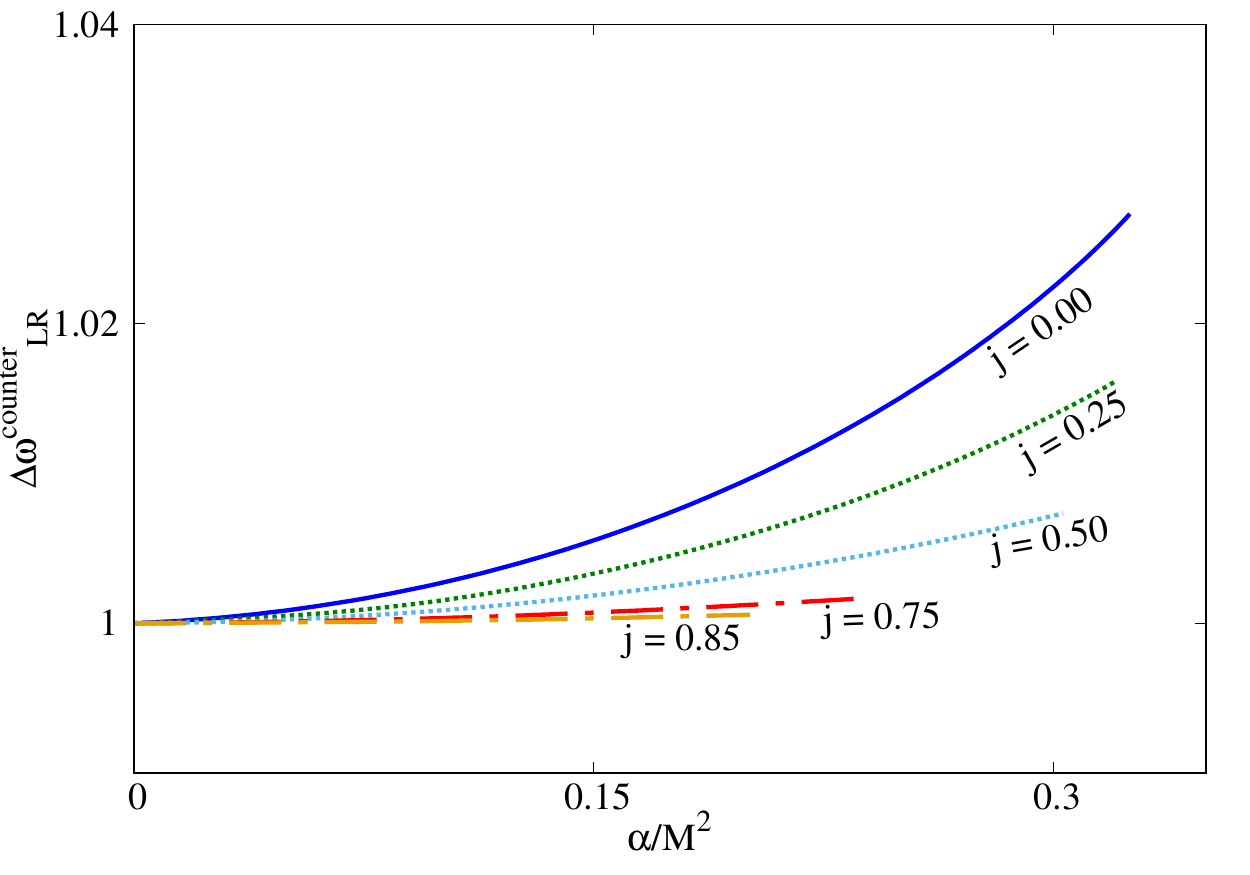}  
\end{center}
  \vspace{-0.5cm}
\caption{ Ratio between the angular frequency at the LR between EsGB BHs and Kerr BHs, for co-rotating orbits (left panel)
and  counter-rotating orbits (right panel).}
\label{Fig:Horndeski_wLR}
\end{figure}

Fig. \ref{Fig:Horndeski_wLR} shows the ratio between the angular frequency at the LR of EsGB BHs and Kerr BHs, for both co-rotating
and counter-rotating orbits, $\Delta \omega_{LR}^{\text{counter}}$, defined in an analogous way to~\eqref{ratios}, 
 with different values of spin, $j$, as a function of the reduced coupling constant, $\alpha/M^2$.  The overall behaviour is very similar to the one discussed above for the ISCO frequency. The main difference for the LR case is that the maximal deviation for both types of orbits is smaller than the corresponding orbits at the ISCO.

 \section{Conclusions and remarks}
\label{sec6}
In this work we have constructed the spinning generalisations of the static 
BHs in the shift symmetric Hordenski model. This is a family of asymptotically flat,  stationary, axially symmetric BHs, that are non-singular on and outside an event horizon. The domain of existence of these solutions is naturally described by two dimensionless parameters: the dimensionless coupling constant of the model, $\alpha/M^2$, and the dimensioness spin of the BHs, $j=J/M^2$. 
Then, the domain of existence is bounded by four special limiting behaviours: the GR limit (when $\alpha=0$), the static limit (when $j=0$), the extremal limit, when the surface gravity of the solutions vanishes, and a critical set of solutions for which a horizon ceases to exist. This last boundary has an important implication. For non-zero $\alpha$ it means there is minimum mass (and hence) size for BHs. Thus there is a mass gap with respect to the Minkowski vacuum, which is also a solution of the theory. 

This non-GR property also occurs for the Einstein-dilaton-GB model discussed  $e.g.$
in~\cite{Kleihaus:2015aje,Kleihaus:2011tg}. Other properties of the BHs we have constructed and analysed in this paper also parallel the solutions found in the Einstein-dilaton-GB model. This similarity of properties was antecipated by the observation made in the introduction:  the linearisation of the action of Einstein-dilaton-GB model 
  \begin{eqnarray}  
\label{actionEGBd}
S=
 \int d^4x \sqrt{-g} \left[R - \frac{1}{2}
 \partial_\mu \phi\partial^\mu \phi
 + \alpha  e^{\phi} R^2_{\rm GB}   \right] \ , 
\end{eqnarray} 
reduces to (\ref{action}) in the limit of small $\phi$, $i.e.$ $e^{\phi}\simeq 1+\phi$, by virtue of~\eqref{totder}.
Since
the scalar field takes rather small values for  typical Einstein-dilaton-GB BHs,
the shift symmetric EsGB BHs
with the same input parameters 
provide  a reasonable approximation - see, for instance, the bottom left panel of Figure~\ref{sol1} for the scalar field magnitude of a typical solution.
Thus, the domain of existence 
of the Einstein-dilaton-GB and EsGB BHs
are indeed quite similar, as confirmed by the results in this work.

Yet, there are both qualitative and quantitative differences between the two models. An intriguing property of the model we have focused on, that does not occur for the Einstein-dilaton-GB model, 
  is the scalar  charge-temperature relation
(\ref{Qs}). In fact, also the Smarr law is different in both models.
Quantitatively, the correspondence between the two models 
holds only for small enough values of $\alpha/M^2$ and $j$.  
 For example, the critical 
value of the ratio $\alpha/M^2$ is
  $0.3253 $
for the spherically symmetric solutions in this work
(being fixed by an algebraic condition between the horizon size and the coupling constant $\alpha$, Eq. (\ref{cond}))
and  $0.1728 $
 for 
Einstein-dilaton-GB BHs (in which case the generalization of (\ref{cond}) includes,  as well,
a dependence on the value of the scalar field at the horizon, see $e.g.$  Ref.\cite{Kanti:1995vq}).
Moreover, a specific feature of  the Einstein-dilaton-GB model is the occurrance, near  the critical configuration
of a small  secondary branch of BH solutions 
\cite{Torii:1996yi,Alexeev:1996vs,Guo:2008hf}.
Along this
branch, the mass increases with decreasing horizon radius.  
This secondary branch appears to be absent in the EsGB case.
 
Finally, let us remark that the way the SZ solutions circumvent the no-scalar-hair theorem also applies to the model herein~\cite{Hui:2012qt}. This occurs by violating the assumption that the current associated to the shift-symmetry should be finite at the horizon. For the static SZ BHs this current diverges on the horizon. This, however, does not induce any physical pathologies. We have checked that this current (squared) diverges at the horizon also in  the spinning BHs reported in this work.

\section*{Acknowledgements}
J. D. is supported by the FCT grant SFRH/BD/130784/2017. This work is supported by the Center for Research and Development
in Mathematics and Applications (CIDMA) through the Portuguese
Foundation for Science and Technology
(FCT - Fundacao para a Ci\^encia e a Tecnologia),
references UIDB/04106/2020 and UIDP/04106/2020 and by national funds (OE), through FCT, I.P., in the scope of the framework contract foreseen in the numbers 4, 5 and 6 of the article 23, of the Decree-Law 57/2016, of August 29,
changed by Law 57/2017, of July 19. We acknowledge support  from the projects PTDC/FIS-OUT/28407/2017 and CERN/FIS-PAR/0027/2019.   This work has further been supported by  the  European  Union's  Horizon  2020  research  and  innovation  (RISE) programme H2020-MSCA-RISE-2017 Grant No.~FunFiCO-777740. The authors would like to acknowledge networking support by the
COST Action CA16104.

\appendix

\setcounter{equation}{0}

 \section{Perturbative solutions}

 \subsection{Spherically symmetric black holes}
\label{a1}
The Schwarzschild BH is not a solution of the model~\eqref{action} with~\eqref{shift-symm}, 
since   $R^2_{\rm GB}\neq 0$.
Nonetheless, one can construct a perturbative solution around it, 
 as a power series in $\beta$ defined in~\eqref{beta}.
Therefore, we consider a generic expansion\footnote{Note that $\phi_1(r)$ is a  nodeless function,
corresponding to the solution of the scalar field eq. (\ref{KG-eq}) in a fixed Schwarzschild background.
Moreover, one can show analytically that the scalar field remains nodeless even with a non-perturbative approach.}
\begin{equation}
\label{pert}
N (r)=\left(1-\frac{r_H}{r }\right)\sum_{k\geqslant 0}  \beta^k h_k(r)\ , \qquad \sigma(r)=\sum_{k\geqslant 0} \beta^k  \sigma_k(r) \ ,
\qquad 
\phi(r)=\sum_{k\geqslant 1} \beta^k \phi_k(r) \ .
\end{equation}
 The horizon is still located at $r=r_H$. Then, one solves the EsGB equations 
 order by order in $\beta$.

The choice~\eqref{pert} leads to a particularly simple structure of the equations for the functions
$\{h_k (r)$, $\sigma_k(r)$, $\phi_k (r)\}$,
which can   easily be solved to an arbitrary order. We have done it up to $k=12$.
These functions are polynomials in $x=r_H/r$,
the expression of the first few terms being
 \begin{eqnarray}
\nonumber
&&
h_0(r)=1,~~h_1(r)=0,~~
h_2(r)=
-\frac{49}{5 }x
-\frac{29}{5 }x^2
-\frac{19}{5 }x^3
+\frac{203}{15 }x^4
+\frac{218}{15 }x^5
+\frac{46 }{3 }x^6,
\\
\label{perxt}
&&
\sigma_1(r)=0,~~
\sigma_2(r)=
-\left(
2x^2
 +\frac{8 x^{3}}{3 }
+7 x^{4}
 +\frac{32 x^{5} }{ 5}
+6x^{6} 
\right),
\\
\nonumber
&&
\phi_1(r)=4x +2x^2+\frac{4x^3}{3 },~~\phi_2(r)=0,
\\
\nonumber
&&
\phi_3(r)=
-\frac{4}{15 } x
+\frac{292 }{15 }x^2
+\frac{1052 }{45 }x^3
+\frac{22 }{5 }x^4
-\frac{476 }{75  }x^5
-\frac{656 }{45   }x^6
+\frac{20  }{3  }x^7
+\frac{58  }{5 }x^8
+\frac{424 }{27 }x^9~.
\end{eqnarray}
Unfortunately, no general pattern can be found and the coefficients of the
terms in the polynomial expressions of  
 $\{h_k(r)$, $\sigma_k(r)$, $\phi_k(r) \}$
become increasingly complicated, with higher powers of $x=r_H/r$.

The corresponding expression of the mass function $m(r)$
follows directly from (\ref{pert})  (we recall that $N=1-2m(r)/r)$.
While $m(r)$ is strictly positive,
its derivative becomes negative in a region close to the event horizon,
the lowest order term being
\begin{eqnarray}
\label{pertm}
m'= 
\left( 
 x^2
+ x^3
+13x^4
+ x^5
+ x^6
-23 x^7 
\right)2\beta^2+\dots \ .
\end{eqnarray}
Thus, for any $\beta$,  one finds 
$m'<0$ for $ r_H \leqslant r \leqslant 1.1049 r_H$.
This implies the existence of
  {\it negative effective energy densities} 
	($i.e.$ $\rho^{\rm (eff)}=-T_t^t<0$)
in the model, a feature confirmed by numerics.

We also display the expression of the first few terms for 
several quantities of interest  
\begin{eqnarray}
&&
M=M^{(0)}
\left(
1+
\frac{49  }{5}\beta^2
+\frac{408253 }{3850  }\beta^4
+\frac{75242913669527}{26533757250  }\beta^6
\right)+\dots,~~
\\
\nonumber
&&
T_H=T_H^{(0)}
\left(
1
- \frac{1}{15  }\beta^2
- \frac{118549}{4950  }\beta^4
- \frac{35399108806973}{26533757250 }\beta^6
\right)+\dots,
\\
\nonumber
&&
S=S^{(0)}
\left(
1
+ \frac{88}{3}\beta^2
+ \frac{162064}{675 }\beta^4
 +\frac{955514545484 }{156080925}\beta^6
\right)+\dots,
\\
\nonumber
&&
 ~Q_s=r_H
\left(
4\beta
-\frac{4}{15}\beta^3
-\frac{237098}{2475}\beta^5
-\frac{70798217613946 }{13266878625}\beta^7
\right)+\dots~,
\end{eqnarray}
with 
$M^{(0)}=\frac{r_H}{2}$, 
$S^{(0)}=\pi r_H^2$,
$T_H^{(0)}=\frac{1}{4\pi r_H}$
the corresponding quantities for the Schwarzschild solution.
One can easily verify that the perturbative expansion
satisfies, order by order, the Smarr relation and the 1st law.

 \subsection{Slowly rotating black holes }
The equations of motion possess a simple solution
for the case of slowly rotating BH  solutions.
The latter 
have been investigated in other gravity theories
(see $e.g.$
\cite{Horne:1992zy,Shiraishi:1992np,Campbell:1990ai})
and usually give an idea about some properties
of the non-perturbative (in the spin parameter) configurations.

To consider slowly rotating BHs 
we assume a metric of the following form
 \begin{eqnarray}
 ds^2=-N(r) \sigma^2(r) dt^2+\frac{dr^2}{N(r)}+r^2\left[d\theta^2+\sin^2 \theta (d\varphi-  W(r) dt) ^2\right] \ , 
\label{sr1}
\end{eqnarray}
with a small
 $W(r) $, such that, to first order in $W$, the above line element takes the (more) familiar form
 \begin{eqnarray}
 ds^2=-N(r) \sigma^2(r) dt^2+\frac{dr^2}{N(r)}+r^2\left(d\theta^2+\sin^2 \theta d\varphi  ^2\right)-2 r^2 \sin^2 \theta W(r) d \varphi dt\ .
\label{sr1w}
\end{eqnarray}
The limit $W(r) =0$
corresponds to the static EsGB BHs discussed above. 
Then it is straightforward to prove that, for small rotation, the
EsGB equations possess the following first integral
 \begin{eqnarray}
  \left\{
	r^3\left[r -4\alpha N(r) \phi'(r)\right] \frac{W'}{\sigma}
	\right\}'=0 \ .
\label{sr2}
\end{eqnarray}
The constant of integration is proportional to 
$J$ -- the angular momentum.
In the absence of a closed form general expression for the EsGB BHs,
the best one can do it to replace in  (\ref{sr2})
the corresponding form of the perturbative solution in $\beta=\alpha/r_H^2$
derived above, and integrate for $W(r)$.
Then a general expression of the form
\begin{eqnarray}
W(r)=\frac{2J}{r^3}\sum_{k\geqslant 0} w_k(r)\beta^{2k} \ ,
\end{eqnarray}
emerges.
 All functions $w_k(r)$ above can be expressed as polynomials in $x=r_H/r$;
here we display the 
  first two functions functions only,
  \begin{eqnarray}
w_0(r)=1\ , \qquad 
w_1(r)=-\left (
\frac{6}{5 }x^2
+\frac{28}{3 }x^3
+3x^4
+\frac{12 }{5 }x^5
-\frac{10 }{3 } x^6
\right) \ .
\end{eqnarray}

This approach holds for the first order in $W$,
thus for an infinitesimally small angular momentum.
Then, physical quantities such as the mass and event horizon area 
 do not change as compared to the static case.
On the other hand, the BH acquires a non-trivial angular momentum 
horizon angular velocity, 
with leading terms
\begin{eqnarray}
\Omega_H=\frac{2J}{r_H^3}
\left(
1
-\frac{63 }{5 }\beta^2
-\frac{206249189  }{1351350 }\beta^4
\right) \ .
\end{eqnarray} 
The corresponding expression of the reduced horizon angular velocity $\omega_H$
is also of interest, with
\begin{eqnarray}
\omega_H= \Omega_H M=
\frac{J}{r_H^2} \left(
1
-\frac{14 }{5 }\beta^2
-\frac{16415506  }{96525  }\beta^4
\right) ~,
%
\end{eqnarray}
a relation which can also be expressed 
in terms of the dimensionless parameters 
$j=J/M^2$ and $\alpha/M^2$
as
 \begin{eqnarray}
\omega_H= 
\frac{j}{4} 
\left[
1
+\frac{21 }{20 }\left( \frac{\alpha}{M^2} \right)^2
+\frac{11390263}{3931200} \left( \frac{\alpha}{M^2} \right)^4
\right]  ~.
%
\end{eqnarray}
Therefore, to these orders in perturbation theory, 
the reduced horizon angular velocity 
{\it increases}
as compared to a similar (slowly rotating) Kerr BH
with the same mass $M$ and angular momentum $J$,
a prediction which agrees with our numerical 
results (see also Figure \ref{Fig:Horndeski_deltaOmegaH}).

 \section{The attractors and the issue of extremal solutions}
\label{apb}
The numerical results suggest that,
unlike the extremal Kerr solution, the extremal EsGB solutions
are not regular.  Evidence for this conjecture is obtained as follows. Instead of solving the full bulk equations searching for extremal solutions,
 one tackles the
construction of the corresponding 
near-horizon configurations.
In this case, one has to solve a co-dimension one problem (the radial dependence being factorized), 
whose solutions are easier to study.

Since this problem was already 
considered in a more general context~\cite{Chen:2018jed} 
(see also the corresponding Einstein-dilaton-GB computation in 
\cite{Kleihaus:2015aje}),
in what follows we shall review the basic results only.
The  idea to consider a  
construction of the near-horizon limit of the extremal rotating BH 
 as a power series in $\alpha$.
The
background solution is taken to be the vacuum near horizon extremal Kerr (NHEK) solution in pure Einstein gravity~\cite{Bardeen:1999px}. 
As we shall see, the  $\alpha^2$-corrections to this solution
are singular and destroy its smoothness.

Following the usual ansatz in the literature (see $e.g.$ \cite{Astefanesei:2006dd}) we consider the following line element
\begin{eqnarray}
\label{metrica}
 ds^2=v_1(\theta) \left ( -r^2 dt^2+\frac{dr^2}{r^2}+\beta^2 d\theta^2 \right)
 +v_2(\theta) \sin^2\theta \left(d\phi+ K r dt \right)^2~,
\end{eqnarray}
 where $0\leqslant r<\infty$, $0\leqslant \theta \leqslant \pi $,  and $\beta,~K$
 are real parameters, while the scalar field depends on $\theta$ only,
\begin{eqnarray}
\label{s1s}
\phi=\phi(\theta) \ .
\end{eqnarray}
%
Also, it  is  convenient to define 
\begin{eqnarray}
\label{u}
\cos\theta=u \ , 
\end{eqnarray}
such that the line element (\ref{metrica}) becomes
\begin{eqnarray}
\label{metric}
 ds^2=v_1(u) \left ( -r^2 dt^2+\frac{dr^2}{r^2}+\beta^2 \frac{du^2}{1-u^2} \right)
 +v_2(u) (1-u^2)\left(d\phi+ K r dt \right)^2 \ .
\end{eqnarray} 

 The functions $v_1(u)$, $v_2(u)$ together with the constants $K,\beta$
satisfy a complicated set of ordinary differential equations which result from the EsGB equations.
These equations (with $\alpha \neq 0 $) appear to possess 
no analytical  solutions.
An approximate solution can be constructed, however, by
  considering an expansion\footnote{More rigorously, 
this expansion is in the dimensionless
parameter $\alpha/J$.} 
in $\alpha$
around the Einstein gravity solution, with
\begin{equation}
\label{expansion1}
v_1(u)=v_{10}(u)+\alpha^2 v_{12}(u)+\dots,~~v_2(u)=v_{20}(u)+\alpha^2 v_{22}(u)+\dots,~~
\phi(u)=\phi_0+ \alpha \phi_1(u)+\dots,
\end{equation}
and
\begin{eqnarray}
\label{expansion2}
K=K_0+\alpha^2 K_{2}+\dots,~~\beta=\beta_0+\alpha^2 \beta_{2}+\dots \ .
\end{eqnarray}
The lowest order terms in the above expansion corresponds to
the Einstein gravity solution  \cite{Bardeen:1999px} 
 \begin{eqnarray}
\label{solE}
 K_0=\beta_0=1 \ , \qquad v_{10}(u)=\frac{J}{16\pi}(1+u^2)\ , \qquad  v_{20}(u)= \frac{J}{4 \pi} \frac{ 1}{(1+ u^2)}\ ,
 \end{eqnarray}   
while $\phi_0$ can be set to zero without any loss of generality.

In the next step, we find 
the expression of $\phi_1(u)$ by solving the eq. (\ref{KG-eq})
in the NHEK background
(\ref {solE})
 \begin{eqnarray}
\label{pert1}
 \frac{d}{du}\left( (1-u^2)\frac{d\phi_1(u)}{du} \right)+\frac{J(1+u^2)}{16 \pi}L_{GB}^{(NHEK)}=0\ ,
 \end{eqnarray} 
$ L_{GB}^{(NHEK)}$ being the GB invariant evaluated for the NHEK geometry,
 \begin{eqnarray}
\label{pert2}
 L_{GB}^{(NHEK)}=-\frac{12288 \pi^2 (-1+15 u^2-15 u^4+u^6)}{J^2 (1+u^2)^6}~.
 \end{eqnarray} 
The general solution of the equation (\ref{pert1}) reads
 \begin{eqnarray}
\label{phi1}
\phi_1(u)=s_0+s_1  \log \left(\frac{1+u}{1-u}\right)
+\frac{32 \pi}{J}
\left[
\frac{2(1-4 u^2-u^4)}{(1+u^2)^3}
+\log \left(\frac{1+u^2}{1-u^2}\right)
\right] \ ,
 \end{eqnarray} 
with $s_0$, $s_1$
arbitrary constants. One can set $s_0=0$ without any loss of generality. 
 For any choice of  $s_1 $ the
function $\phi_1(u)$ necessarily diverges at $u=1$ and/or $u = -1$.
In our approach, we take
 \begin{eqnarray}
s_1=\frac{32 \pi}{J} \ ,
 \end{eqnarray} 
such that $\phi_1(u)$ is divergent at $u=1$ only.

In the next step, we solve for the corrections to the geometry 
as encoded in the functions $v_{12}(u)$ and $v_{22}(u)$.
Since 
the equations for these functions are sourced by a divergent scalar field
 $\phi_1(u)$,
one expects them to be divergent as well.
This is indeed confirmed by our results, 
and one finds
 \begin{eqnarray}
\label{v12}
&&
v_{12}(u)=\frac{\pi}{J}
\bigg(
{\cal F}_1(u)
-32 (4+u(3+4u))\arctan u
-64 (1-u^2) \log (1+u^2)
\\
\nonumber
&&
-\beta_2\frac{J^2}{24 \pi^2}(2+2u^2 +3u \sqrt{1-u^2}\arccos u)
-128 (1-u)^2 \log (1-u)
-64 \sqrt{2} u \sqrt{1-u^2}
\\
\nonumber
&&
+\frac{969}{\sqrt{2}}u\sqrt{1-u^2}\arctan \frac{\sqrt{2}u}{\sqrt{1-u^2}}
-192  \sqrt{2} u \sqrt{1-u^2} \arctanh \frac{\sqrt{1-u^2} }{\sqrt{2}}
\\
\nonumber
&&
+u z_1+(1+(u-4)u)z_2-u \sqrt{1-u^2}z_3
\bigg) \ ,
 \end{eqnarray} 
where $z_1$, $z_2$, $z_3$
are arbitrary constants and we define
 \begin{eqnarray}
\nonumber
&&	
{\cal F}_1(u)=\frac{1}{105 (1+u^2)^5}
\bigg(
 88054 + 26880  u + 759219  u^2 + 161280 u^3 + 1133035u^4 + 376320 
      u^5
	\\
\nonumber
&&		
			+ 1617566u^6 + 430080 u^7 + 1109548 u^8 + 241920 
      u^9 + 377967 u^{10} + 53760 u^{11} + 49331 u^{12} 
\bigg) \ .
 \end{eqnarray} 
A very similar expression
is found for $v_{22}(u)$, with
 \begin{eqnarray}
\label{v22}
&&
v_{22}(u)=\frac{8\pi}{J(1+u^2)^2}
\bigg(
{\cal F}_2(u)
-16 (4-u(3-4u))\arctan u
-32 (1+u)^2 \log (1+u^2)
\\
\nonumber
&&
+\beta_2\frac{J^2}{48 \pi^2}(-5(1+u^2) +\frac{6 u \arccos u}{ \sqrt{1-u^2}} )
-64 (1-u)^2 \log (1-u)
+64 \sqrt{2} \frac{u}{\sqrt{1-u^2}}
\\
\nonumber
&&
+\frac{969}{\sqrt{2}}\frac{u}{\sqrt{1-u^2}}\arctan \frac{\sqrt{2}u}{\sqrt{1-u^2}}
+\frac{192  \sqrt{2} u}{ \sqrt{1-u^2} }\arctanh \frac{\sqrt{1-u^2} }{\sqrt{2}}
\\
\nonumber
&&
-\frac{1}{2}u z_1+\frac{1}{2}(1+4 u +u^2 )z_2-\frac{u}{\sqrt{1-u^2}}  z_3
\bigg) \ ,
 \end{eqnarray} 
with
 \begin{eqnarray}
\nonumber
&&	
{\cal F}_2(u)=\frac{1}{105 (1+u^2)^5}
\bigg(
 40667 - 20160 u + 4707u^2 - 114240  u^3 + 515365 u^4 - 
    255360 u^5 
		\\
		&&	
		\nonumber
		+ 474470  u^6 - 282240 u^7 + 372733 u^8 - 
    154560 u^9 + 174351 u^{10} - 33600 u^{11} + 35035 u^{12}
\bigg) \ .
 \end{eqnarray} 

Then, with the above expressions,
one can prove the existence of
a singularity at the poles of the horizon,
with the Ricci scalar diverging  at $\theta=0$ ($i.e.$ $u=1$)
  \begin{eqnarray}
\label{sing}
 R=-\frac{32768 \pi^3 \alpha^2}{J^3(u-1)}+\frac{49152  \pi^3 \alpha^2}{J^3 }+\mathcal{O}(u-1) \ .
 \end{eqnarray}

Finally, let us mention that
the above perturbative result does not exclude the existence
of regular solutions for $\alpha$ large enough.
Thus we have also attempted to solve  the field equations of the model within a nonperturbative approach,
by  solving a boundary value problem.
The imposed boundary conditions assure the regularity of the configurations at $u=\pm 1$.
However, no  such solutions could be found.

 \begin{small}
 
 \end{small}

 \end{document}